\documentclass[useAMS,usenatbib]{mn2e}
\usepackage{graphicx}
\usepackage{float}
\usepackage{float}
\usepackage{color}
\def\kms{km\,s$^{-1}$}
\def\ene{erg\,s$^{-1}$}

\def\LL{$\lambda\lambda$}
\def\Ha{H$\alpha$}

\def\L{$\lambda$}
\def\BaII{Ba\,{\sc ii}} 
\def\CaII{Ca\,{\sc ii}}
\def\CI{C\,{\sc i}}

\def\CoII{Co\,{\sc ii}} 
\def\CrII{Cr\,{\sc ii}} 
\def\CI{C\,{\sc i}}
\def\OI{O\,{\sc i}}
\def\OII{O\,{\sc ii}}

\def\SiII{Si\,{\sc ii}}
\def\TiII{Ti\,{\sc ii}}
\def\SII{S\,{\sc ii}}
\def\ScII{Sc\,{\sc ii}}
\def\HeI{He\,{\sc i}}
\def\FeII{Fe\,{\sc ii}}

\def\MgII{Mg\,{\sc ii}}
\def\MgI{Mg\,{\sc i}}
\def\NaID{Na\,{\sc i}~D}
\def\NaI{Na\,{\sc i}}

\def\msun{M$_{\odot}$}
\def\lesssim{\mathrel{\hbox{\rlap{\hbox{\lower4pt\hbox{$\sim$}}}\hbox{$<$}}}}
\def\gtrsim{\mathrel{\hbox{\rlap{\hbox{\lower4pt\hbox{$\sim$}}}\hbox{$>$}}}}

\title[SN~2012hn]
{PESSTO monitoring of SN~2012hn: further heterogeneity among faint type I supernovae\thanks{This 
paper is based on observations obtained during the first run of the Public ESO Spectroscopic Survey 
for Transient Objects. Data from the following telescopes are included: NTT(184.D-1140,188.D-3003),
VLT-UT1(089.D-0270), PROMPT, TRAPPIST, DUPONT, MAGELLAN }}
\author[Valenti et al.]{S. Valenti$^{1,2}$
\thanks{e--mail: svalenti@lcogt.net},
F. Yuan$^{3,4}$, S. Taubenberger$^{5}$, K. Maguire$^{6}$, A. Pastorello$^{7}$, 
\and
S. Benetti$^{7}$, S.~J. Smartt$^{8}$, E. Cappellaro$^{7}$ , D.~A. Howell$^{1,2}$, L. Bildsten$^{2,9}$, 
\and
K.  Moore$^{2}$, M. Stritzinger$^{10}$, J.~P. Anderson$^{11}$, S. Benitez-Herrera$^{5}$,  
\and
F. Bufano$^{12}$, S. Gonzalez-Gaitan$^{11}$, M.~G. McCrum$^{8}$, G. Pignata$^{12}$, M. Fraser$^{8}$,  
\and
A. Gal-Yam$^{13}$,  L. Le Guillou$^{14}$, C. Inserra$^{8}$, D.~E. Reichart$^{15}$, R. Scalzo$^{3}$,
\and 
M. Sullivan$^{16}$, O. Yaron$^{13}$, D. R. Young$^{8}$\\
$^{1 ~}$ Las Cumbres Observatory Global Telescope Network, 6740 Cortona Dr., Suite 102, Goleta, CA 93117, USA \\
$^{2 ~}$ Department of Physics, University of California, Santa Barbara, Broida Hall, Mail Code 9530, Santa Barbara, CA 93106-9530, USA) \\
$^{3 ~}$ Research School of Astronomy and Astrophysics, The Australian National University, Weston Creek, ACT 2611, Australia \\
$^{4 ~}$ ARC Centre of Excellence for All-sky Astrophysics (CAASTRO) \\
$^{5 ~}$ Max-Planck-Institut f\"{u}r Astrophysik, Karl-Schwarzschild-Str. 1, 85741 Garching bei M\"{u}nchen, Germany) \\
$^{6 ~}$ Department of Physics (Astrophysics), University of Oxford, DWB, Keble Road, Oxford, OX1 3RH, UK\\
$^{7 ~}$ INAF Osservatorio Astronomico di Padova, Vicolo dell'Osservatorio 5, 35122 Padova, Italy \\
$^{8 ~}$ Astrophysics Research Centre,  School of Mathematics and Physics, Queens University Belfast, Belfast BT7 1NN, UK \\
$^{9 ~}$ Kavli Institute for Theoretical Physics, Kohn Hall, University of California, Santa Barbara, CA 93106-4030, USA\\
$^{10}$ Department of Physics and Astronomy, Aarhus University, Ny Munkegade 120, DK-8000 Aarhus C, Denmark \\
$^{11}$ Departamento de Astronomia, Universidad de Chile, Casilla 36-D, Santiago, Chile\\ 
$^{12}$ Departamento de Ciencias Fisicas, Universidad Andres Bello, Avda. Republica 252, Santiago, Chile \\
$^{13}$ Department of Particle Physics and Astrophysics, The Weizmann Institute of Science, Rehovot 76100, Israel \\
$^{14}$ UPMC Univ. Paris 06, UMR 7585, Laboratoire de Physique Nucleaire et des Hautes Energies (LPNHE), 75005 Paris, France \\
$^{15}$ University of North Carolina at Chapel Hill, Campus Box 3255, Chapel Hill, NC 27599-3255, USA\\
$^{16}$ School of Physics and Astronomy, University of Southampton, Southampton, SO17 1BJ, UK
}

\begin{document}

\date{Accepted .....; Received ....; in original form ....}


\maketitle

\begin{abstract}
We present optical and infrared monitoring data of SN~2012hn collected
by the Public ESO Spectroscopic Survey for Transient Objects (PESSTO).
We show that SN~2012hn has a faint peak magnitude (M$_R$\,$\sim$\,$-15.65$) 
and shows no hydrogen and no clear evidence for helium in its spectral 
evolution.
Instead, we detect prominent \CaII{} lines at all epochs, which relates 
this transient to previously described `Ca-rich' or `gap' transients. 
However, the photospheric spectra (from $-3$ to +32\,d with respect to 
peak) of SN~2012hn show a series of absorption lines which are unique, 
and a red continuum that is likely intrinsic rather than due to extinction. 
Lines of \TiII{} and \CrII{} are visible. This may be a temperature effect, 
which could also explain the red photospheric colour. A nebular spectrum 
at +150\,d shows prominent \CaII, \OI, \CI{} and possibly \MgI{} lines
which appear similar in strength to those displayed by core-collapse SNe. 
To add to the puzzle,
SN~2012hn is located at a projected distance of 6 kpc from an E/S0 host 
and is not close to any obvious starforming region. Overall SN~2012hn 
resembles a group of faint H-poor SNe that have been discovered recently 
and for which a convincing and consistent physical explanation is still 
missing. They all appear to  explode preferentially in remote locations 
offset from a massive host galaxy with deep limits on any dwarf host 
galaxies, favouring old progenitor systems. SN~2012hn adds heterogeneity 
to this sample of objects. We discuss potential explosion channels 
including He-shell detonations and double detonations of white dwarfs as 
well as peculiar core-collapse SNe. 
\end{abstract}

\begin{keywords}
supernovae: general -- supernovae: SN~2012hn --  galaxies: NGC~2272 
\end{keywords}

\section{Introduction}
\label{parintroduction}

In recent years,  an increasing number of new transients with unusual 
properties have been discovered thanks to the advent of modern,  wide-field 
optical transient surveys, such as the Texas Supernova Search 
\citep[TSS, ][]{QuimbyRobertMichael2006} the Catalina Real-Time Transient Survey 
(CRTS, Drake et al. 2009), the Palomar Transient Factory \citep[PTF, ][]{2009PASP..121.1334R}
and the Panoramic Survey Telescope and Rapid Response System 
\citep[Pan-STARRS, ][]{Kaiser2002}. Many new objects  have been found in regions of 
the time-scale/luminosity parameter space unexplored before (e.g. faint and/or fast 
evolving transients) or were located in regions of the sky undersampled by 
traditional \emph{targeted}-surveys (transients in faint hosts or in the outskirts 
of bright galaxies).

Driven by their observed characteristics, attempts have been made to arrange these
novel transients in new classes, trying also to revise the classification of historical 
SNe in the context of non-standard scenarios. In particular, major efforts have been 
made to theoretically explain new transients that may not  be due to the canonical 
mechanisms of iron core collapse or the thermonuclear explosion of 
near-Chandrasekhar-mass white dwarfs.

Among these new transients, a wide group of peculiar hydrogen-free 
supernovae (SNe) has recently been studied in detail 
(including SN~2008ha: \citealt{2009Natur.459..674V,2009AJ....138..376F};      
SN~2005E: \citealt{2010Natur.465..322P};  
SN~2002bj: \citealt{2010Sci...327...58P};          
SN~2010X:  \citealt{2010ApJ...723L..98K};
SN~2005cz: \citealt{2010Natur.465..326K}; 
PTF09dav: \citealt{2011ApJ...732..118S,2012ApJ...755..161K};  
SN~2010et/PTF10iuv\footnote{SN~2010et  was  discovered by
\protect\cite{2010CBET.2339....1D} and independently, but never
announced, by the Palomar Transient Factory (PTF, 
\citealt{2009PASP..121.1334R}).}: \citealt{2012ApJ...755..161K}).

SNe~2002bj and 2010X are two rapidly declining SNe 
\citep{2010ApJ...723L..98K}. However, beyond the fast evolution 
they have in common, they differ in absolute magnitude 
and photospheric velocities.
If the light curve is powered by $^{56}$Ni decay, their fast evolution 
indicates a very small ejected mass 
\citep[e.g.  0.16 \msun{} for SN~2010X, ][]{2010ApJ...723L..98K}.
Accretion-induced collapse of an O-Ne-Mg white dwarf   
\citep{Nomoto1984,2009arXiv0908.1127M},
a thermonuclear helium shell detonation on a white dwarf  \citep{2007ApJ...662L..95B} 
and  fallback core collapse \citep{2010ApJ...719.1445M} are the most common scenarios 
proposed to explain these two SNe.
 
SN~2008ha is probably  the faintest hydrogen poor SN ever studied so far. Its extremely 
low photospheric velocity and fast light-curve evolution (but not as fast as in the rapidly
declining SNe mentioned above) are consistent with a very low mass and kinetic energy  
\citep{2009Natur.459..674V}. Given these characteristics, the same models as mentioned 
above have been proposed to explain SN~2008ha. However, a few additional models have 
been suggested, including Fe or ONe core-collapse 
triggered by electron captures \citep{2009ApJ...705L.138P} or a  failed deflagration of a carbon 
oxygen white dwarf \citep{2009AJ....138..376F}.
Recently, due to its spectral similarity with SN 2002cx, SN 2008ha has been included 
as a low-velocity and low-mass member of the SN~Iax subclass \citep{Foley2013}.
SNe~Iax include objects with spectra similar to those of SN~2002cx, but that show a wide 
range of line velocities and luminosities \citep{Foley2013}. Most SNe belonging to this class 
exploded in late-type galaxies.

In this paper, we will mostly focus on a group of objects often referred to as 
`Ca-rich transients' \citep{2003IAUC.8159....2F,2010Natur.465..322P}. 
These are faint type I SNe, which reach peak absolute 
magnitudes of $M_{R} \gtrsim -16$ and exhibit photospheric expansion velocities
of $\sim$\,11000 \kms. They preferentially occur in the outskirts of their presumed 
host galaxies \citep{2010Natur.465..322P,2012ApJ...755..161K} and their designation 
stems from prevalent [\CaII{}] lines that dominate their spectra soon after maximum.

The first recognized member of this class is SN 2005E, for which 
\cite{2010Natur.465..322P} inferred a calcium mass of 0.135 \msun, 
which is much larger than ever seen in other SNe, combined with a very low 
total ejected mass ($\sim$ 0.3 \msun). Its location, the He-rich spectra, 
the low luminosity, and the small ejected mass led \cite{2010Natur.465..322P} 
to suggest a helium detonation on a helium-accreting white dwarf.
Other members of this group of objects are SN~2005cz, SN~2010et 
and PTF11bij. They all show strong Ca lines, He lines in early spectra,
weak O features in late spectra, a modest luminosity and a fast transition 
to the nebular phase. The latter property is a robust evidence of a small ejecta mass.

\begin{table*}
 \centering
  \begin{minipage}{140mm}
  \caption{Journal of spectroscopic observations. The spectra are available in 
  electronic format on WISeREP (the Weizmann interactive supernova data repository 
  - \protect\citealt{2012PASP..124..668Y}). }
  \label{tabseqspec}  
  \begin{tabular}{@{}cccccc@{}}
  \hline   
Date  &   JD  &  Phase\footnote{Relative to $R$-band maximum light (JD = 2,456,034.5).} 
& Range & Resolution FWHM\footnote{FWHM of night-sky emission lines.} 
&  Equipment\footnote{
NTT $=$ ESO New Technology Telescope; 
$~$ VLT $=$ ESO Very Large Telescope;
$~$ MAG $=$ Magellan}
 \\
 &  $-$2,400,000 & (days) & (\AA) & (\AA) & \\
 \hline
2012 Apr 14 & 56031.52  & $-3 $ & 3800--9200    & 27    &  NTT+EFOSC2+Gr13 \\
2012 Apr 20 & 56038.54  & $+4 $ & 3400--10000   & 15    &  NTT+EFOSC2+Gr11/Gr16 \\
2012 Apr 22 & 56040.49  & $+6 $ & 3400--10000   & 15    &  NTT+EFOSC2+Gr11/Gr16 \\
2012 Apr 29 & 56047.50  & $+13$ & 3400--10000   & 15    &  NTT+EFOSC2+Gr11/Gr16 \\
2012 May 11 & 56059.48  & $+25$ & 3000--24000   & 1,2.8 &  VLT+XSHOOTER \\
2012 May 18 & 56066.49  & $+32$ & 3700--9000    &  8    &  MAG+LDSS3+VPH \\
2012 Sep 13 & 56183.87  & $+149$ & 3000--10000  &  1    &  VLT+XSHOOTER \\
2012 Sep 14 & 56184.87  & $+150$ & 3000--10000  &  1    &  VLT+XSHOOTER \\
\hline
\end{tabular}
\end{minipage}
\end{table*}

\begin{figure*}
\begin{center}
  \includegraphics[width=8.5cm,height=7cm]{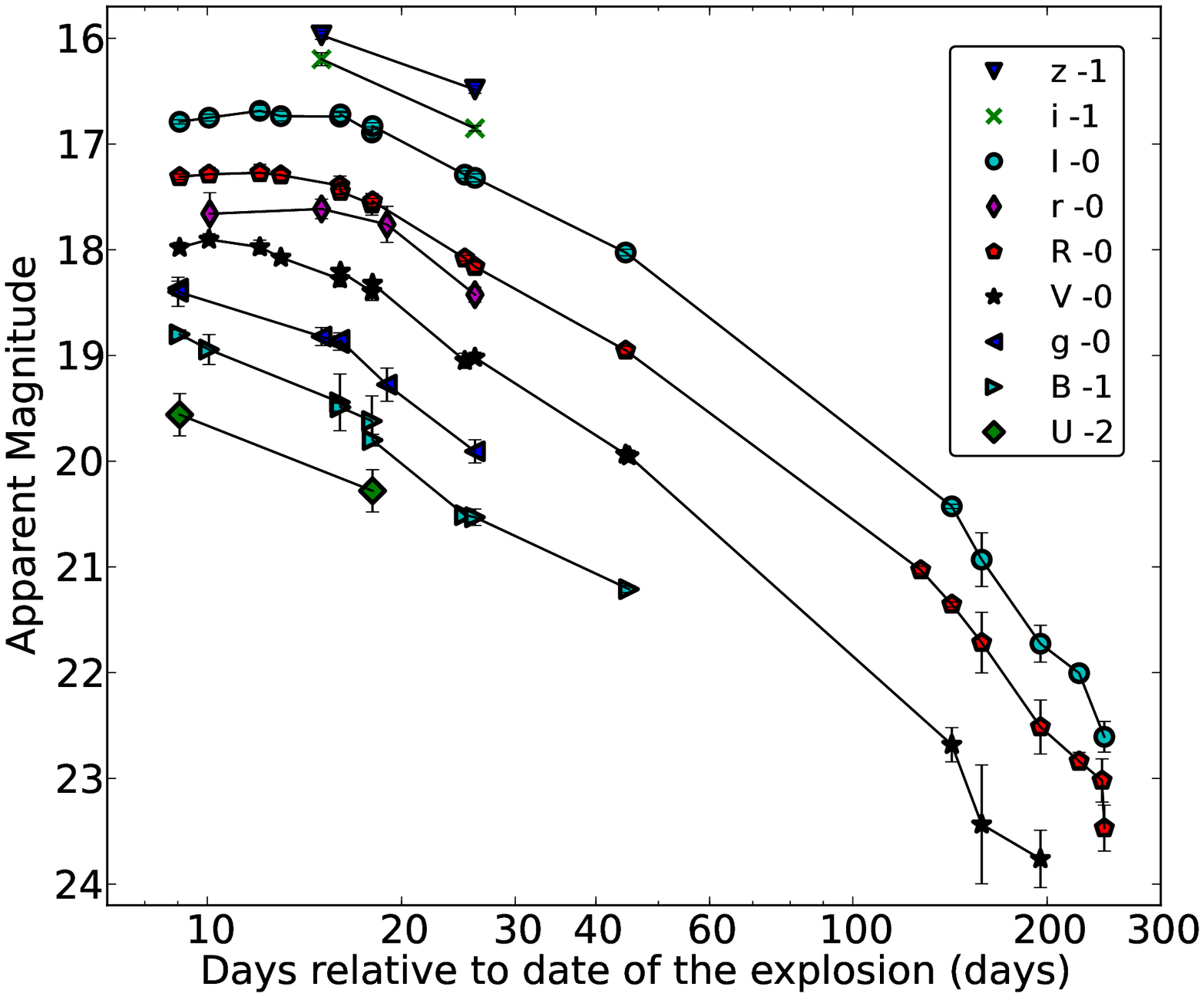}
  \includegraphics[width=8.5cm,height=6.5cm]{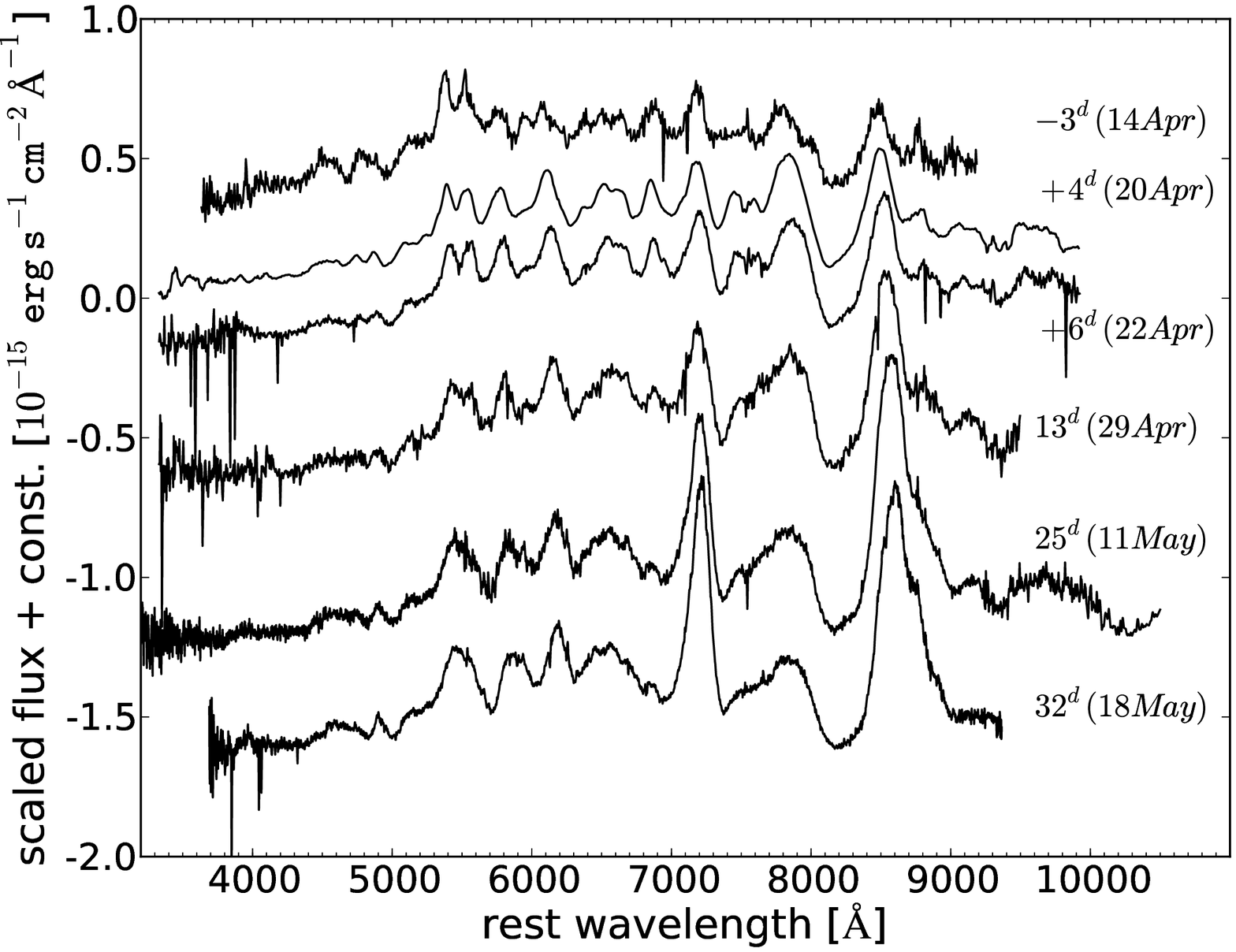}
  \caption{Left: Photometry of SN~2012hn in $UBgVrRiIz$. 
  The date of the explosion has been fixed 
  12 $\pm$ 3 days before $R$-band maximum light ($\rm{JD_{exp}}$ = 2,456,022.5).
  Right: Full sequence of photospheric spectra collected for SN~2012hn.}
  \label{fig:lc}
\end{center}
\end{figure*}

The detection of  strong [\CaII] lines is not a unique property of this 
group of SNe, since these are occasionally detected in other SN types as 
well. For example, faint type IIP SNe also exhibit strong [\CaII] lines at 
late phases \citep{2004MNRAS.347...74P}. Similarly, SN~2008ha shows strong 
[\CaII] lines. However, this SN differed from all other `Ca-rich objects' 
in being very faint, evolving rapidly and having a remarkably low photospheric 
velocity.
Since the detection of strong [\CaII] features may be common to different 
SN types (including objects without an enhanced Ca abundance), we will 
instead refer to these objects as faint type I SNe, based on their 
observational characteristics as H-deficient SNe with $M \gtrsim -16$.

Presented in this paper are observations of the faint Type I SN~2012hn, 
acquired by the Public ESO Spectroscopic Survey for Transient Objects 
(PESSTO\footnote{In response to the first call by the European Southern 
Observatory (ESO) for public spectroscopic surveys, and particularly prompted 
by the opportunities provided by currently running wide-field surveys, PESSTO 
was awarded 90 nights per year on the New Technology Telescope (NTT) for four 
years, with continuation dependent on a mid-term review and the possibility 
for a fifth year of operations depending on the future of La Silla.}).
The overall science goal of PESSTO is to provide a public data base of 
high-quality optical\,+\,near-infrared (NIR) spectral time series of 150 
optical transients covering the full parameter space that modern surveys 
deliver in terms of luminosity, host metallicity and explosion mechanisms. 
SN~2012hn is a PESSTO Key Science target as it conforms with the science 
goal to study unusual and unexplained transients.

The data of SN~2012hn are presented and compared with other faint 
type I SNe: SN~2005E, SN~2005cz, SN~2010et and PTF11bij. 
\cite{2012ApJ...755..161K} included in this class also PTF09dav 
\citep{2011ApJ...732..118S} and SN~2007ke \citep{2007CBET.1101....1F}, and 
identified possible other class members in archival data.

PTF09dav is the only object within this group that spectroscopically resembles 
1991bg-like SNe~Ia, but with scandium and strontium features in photospheric 
spectra \citep{2011ApJ...732..118S}, and weak hydrogen emission during the 
nebular phase \citep{2012ApJ...755..161K}. It was initially identified as a 
peculiar sub-luminous SN~Ia \citep{2011ApJ...732..118S} and later included by
\cite{2012ApJ...755..161K} in the group of `Ca-rich transients'. 
Note that scandium has never been identified in supernovae whose thermonuclear origin is uncontroversial, but that it is frequently observed in SNe II \citep{2004MNRAS.347...74P}
 and occasionally in SNe Ib/c \citep[e.g. SN~2007gr, ][]{2008ApJ...673L.155V}. 
 SN~2007ke has  been added on the basis of a strong feature at 7300\,\AA{} identified as 
[\CaII] \LL{}7291,7323 in a spectrum 19 days after maximum.

Whenever possible, PTF09dav and SN~2007ke have 
been included in the discussion even though their link with the other 
objects is less secure. The other possible 
members of this class, identified in archival data, are not included in this 
paper. The lack of quality data for most of them makes it even more difficult 
to establish their physical link with the group of faint type I SNe.
Some of the SNe Iax recently presented by \cite{Foley2013} are in principle 
faint type I SNe (e.g. SN~2008ha), but their spectra are quite different, 
they explode in a different environment \citep{Lyman2013} and they will not be 
discussed extensively here.

\section{data}
\label{data}

SN~2012hn was  discovered by  the Catalina Real-Time  Transient Survey
(CRTS) on 2012 March 12.43  (UT dates are used throughout this paper)
and classified at the ESO-NTT on 2012 March 14.02  as 
a peculiar SN~Ic with some unidentified features \citep{2012ATel.4047....1B} 
during an ELP\footnote{The European Large Program was a large NTT program 
(PI S. Benetti) to study nucleosyntesis of SNe. The project ended in September 
2012 and shared time with PESSTO during period 89.} observing run. 
No objects with similar spectra were identified in the Padova Supernova archive,
 making SN~2012hn an interesting and rare type of SN.

Given its right ascension (see next section), SN~2012hn could be followed 
only for $\sim$\,2 months after discovery. We collected 4 spectra with
NTT+EFOSC2 in April and two spectra (with VLT+XSHOOTER and Magellan+LDSS3) 
in May. Two nebular spectra were obtained with VLT+XSHOOTER on 2012 September
13.37 and September 14.37. Photometry was collected mostly with NTT+EFOSC2 and the 
PROMPT 5 telescope \citep{2005NCimC..28..767R}, with a handful of points 
added by the TRAPPIST and Du Pont telescopes. The log of our 
spectroscopic observations of SN~2012hn is reported in Table~\ref{tabseqspec}, 
the photometry in Tables~\ref{tablandolt} and \ref{tabsloan}.

The NTT spectra were reduced using a custom-built {\sc python/pyraf}
package developed by the author to reduce PESSTO data. Spectral reduction 
within the NTT pipeline includes corrections for bias and fringing, 
wavelength and flux calibration, correction for telluric absorptions  
and a check on the correctness of the wavelength calibration using the 
atmospheric emission lines.
The VLT+XSHOOTER spectra were reduced using the ESO-XSHOOTER    
pipeline version 1.0.0 under the {\sc gasgano} framework. The Magellan+LDSS3
spectrum was reduced  in a standard fashion using  {\sc iraf}. 
Imaging data were reduced  using the QUBA  pipeline \citep[see
][]{2011MNRAS.416.3138V}.  The SN photometry was measured trough 
PSF fitting and calibrated against a set of local sequence stars. 
The latter were calibrated with respect to Landolt and Sloan standard 
fields during two photometric nights at the NTT.
The magnitude errors account for the uncertainties in the PSF fit and  
the zero points (photometric calibration). The magnitudes of the local  
sequence stars are reported in Tables~\ref{tabseqstar1} (Landolt system) 
and \ref{tabseqstar2} (Sloan system). Our spectral sequence and
light curves are shown in Fig.~\ref{fig:lc}.

Already in the first spectrum, SN~2012hn showed prominent lines
that are usually observed in SNe with relatively low temperatures. The
spectra are red with a drop in luminosity below 5300\,\AA{} that yields 
a SN $B-V$ colour $\sim$\,2. The $B$ band evolves much faster than other 
optical bands, and its peak likely occurred  earlier than the discovery.  
Whether SN~2012hn is intrinsically very red or it is red because of dust 
extinction along the line of sight is not obvious. Unfortunately, for this 
kind of transients a robust method to estimate the extinction does not 
exist. For this reason, in the next section we will discuss different 
reddening scenarios.

\section{Host and reddening}
\label{sec:reddening}

\begin{figure}
\begin{center}
  \includegraphics[width=8.cm,height=6cm]{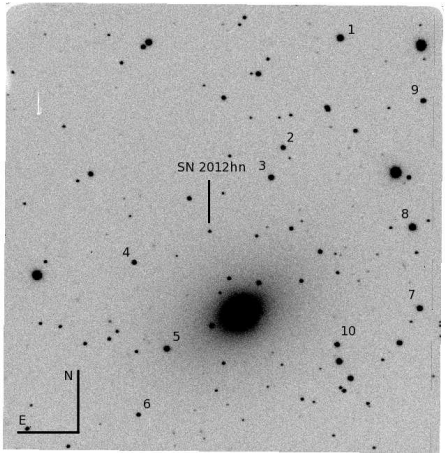}
  \caption{NTT+EFOSC2 $R$-filter image of the field of SN~2012hn 
  ($4\times4$ arcmin$^2$). The local sequence stars of Tables~\ref{tabseqstar1} 
  and \ref{tabseqstar2} are labelled.}
  \label{fig:localstars}
\end{center}
\end{figure}

SN~2012hn is located at $\alpha = 06^\mathrm{h}42^\mathrm{m}42\fs55$
and $\delta = -27\degr26\arcmin49\farcs8$, $44.2$ arcsec north and
$18.7$ arcsec east of the nucleus of NGC~2272, an E/S0-type 
galaxy\footnote{http://leda.univ-lyon1.fr/} 
(see Fig.~\ref{fig:localstars} and Table~\ref{tabsummary}).
Assuming a distance of 26.8 $\pm$ 1.9 Mpc for NGC~2272\footnote{From NED, 
including a correction for Local-Group infall onto the Virgo cluster.}, 
SN~2012hn lies at a projected distance of 6.2 kpc from the centre of 
the host. 
To compare the SN location against the host-galaxy light\,/\,stellar-mass 
distribution, we estimate the enclosed light fraction using the following 
method: we model the projected radial surface density profile of the
host galaxy using fluxes extracted in elliptical apertures and calculate 
the ratio of light within the ellipse defined by the SN location and the 
integrated galaxy flux \citep{Yuan2013}. We find that SN~2012hn 
lies at a distance of more than three times the half-light ellipse. The 
ellipse passing through the SN encloses 88\% of its $R$-band light or 
92\% of its $K$-band light (using images from the 2MASS archive). 
The latter is often used as a proxy for stellar mass. 

SN~2012hn represents another example of a peculiar event occurring in 
the outer regions of an early-type host. Although it is not as distant 
as some others \citep[e.g. see ][]{2012ApJ...755..161K}, the 6.2 kpc is 
a projected distance, and the deprojected offset will likely be larger.
Such an apparent preference for remote locations is different from type 
Ia and type II SNe that roughly follow the light distribution in their  
hosts \citep{2008MNRAS.388L..74F}, 
and from SNe Ib/c that are more centrally concentrated in galaxies
\citep{2008ApJ...687.1201K,2009MNRAS.399..559A,2011A&A...530A..95L}.

\begin{table}
\caption{Main parameters for SN~2012hn and its host galaxy.}
\label{tabsummary}
\begin{tabular}{ll}
\hline
Parent galaxy                      & NGC 2272                   \\
Galaxy type                        & E/S0 (LEDA)\,/\,SAB0 (NED) \\
RA (J2000)                         & $06^\mathrm{h}42^\mathrm{m}42\fs55$ \\
Dec (J2000)                        & $-27\degr26\arcmin49\farcs8$ \\
Recession velocity$^a$             & 1917 \kms \\
Distance modulus$^b$               & $32.14 \pm 0.15$ mag \\
$E(B-V)_\mathrm{NGC\ 2272}$         & 0.2 mag \\
$E(B-V)_\mathrm{MW}^c$             & 0.105 mag \\
Offset from nucleus                & $18.75$ arcsec E, $44.20$ arcsec N \\
Maximum epoch in $V$ (JD)          & $2456032.5\pm 1.0$ (Apr 15, 2012) \\
Maximum epoch in $R$ (JD)          & $2456034.5\pm 1.0$ (Apr 17, 2012) \\
\hline
\end{tabular}\\
$^a$ From LEDA, velocity corrected for Local-Group infall onto the Virgo cluster.\\ 
$^b$ From NED, corrected for Local-Group infall onto the Virgo cluster and assuming 
$H_0 = 72$ \kms\,Mpc$^{-1}$.\\
$^c$ \protect\cite{1998ApJ...500..525S}.
\end{table}

\begin{figure}
\begin{center}
   \includegraphics[width=8.5cm,height=7cm]{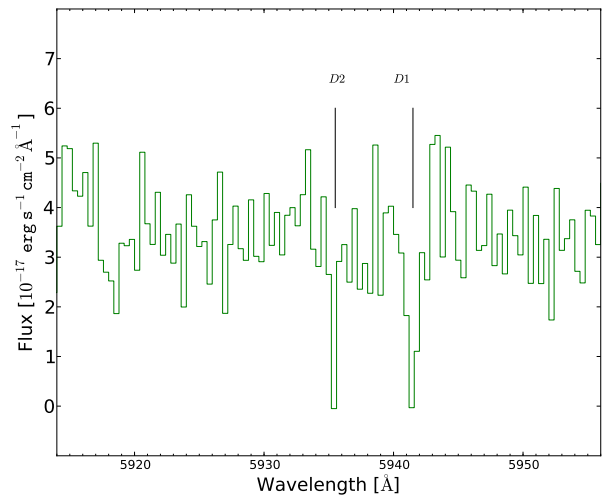}
  \caption{\NaID{} from the host galaxy.}
  \label{fig:naid}
\end{center}
\end{figure}

\begin{figure}
\begin{center}
  \includegraphics[width=9.cm,height=8cm]{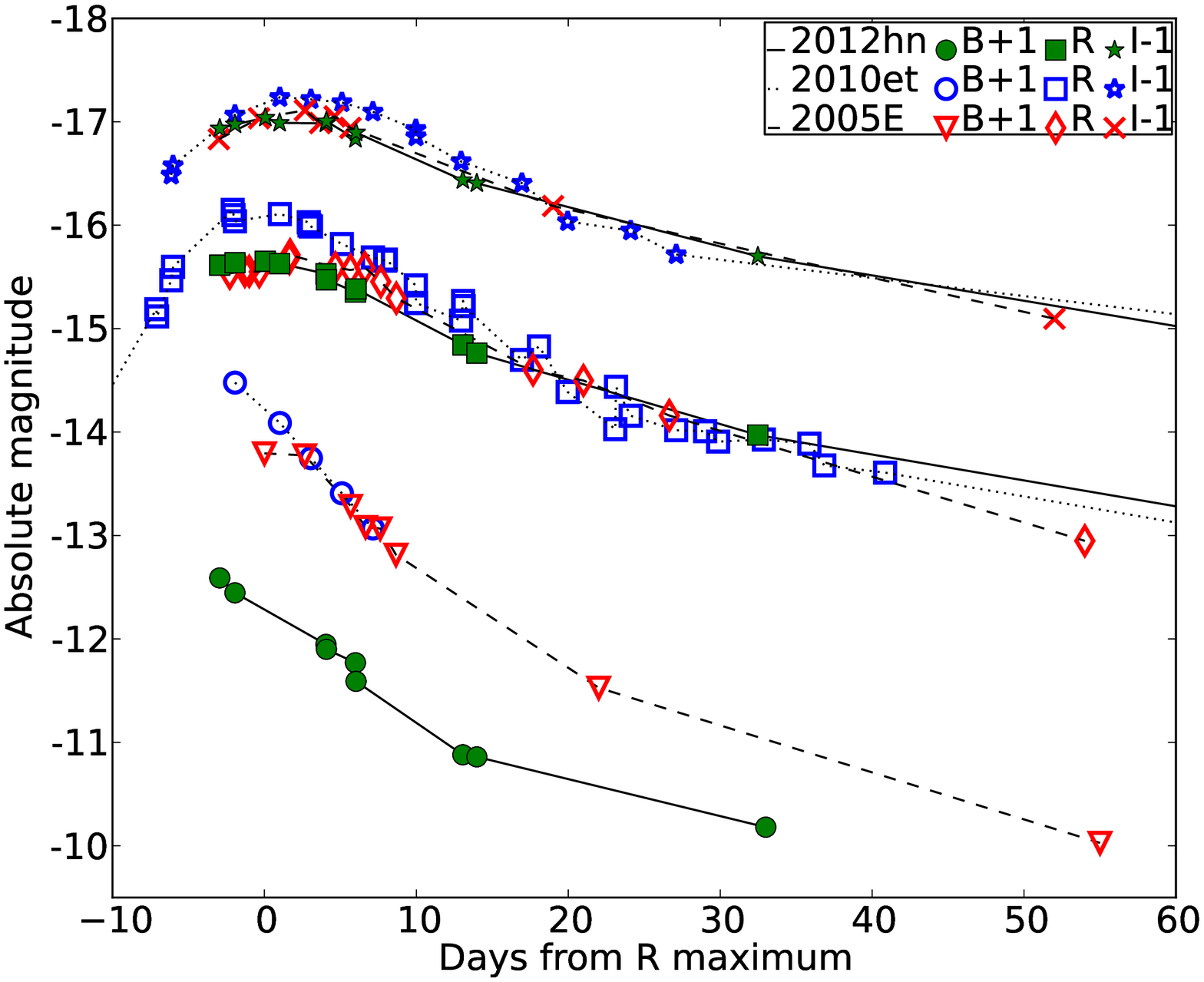}
  \caption{Light-curve comparison of SNe~2005E, 2010et and 2012hn.}
  \label{fig:lccomp}
\end{center}
\end{figure}

The remote location of SN~2012hn in the host galaxy indirect suggests
a  low extinction along the line of sight. 
On the other hand, the observed colours of SN~2012hn are rather extreme. 
Few (if any) SNe have been observed to be so red, which might therefore 
suggest at least some extinction.
A comparison of the SED with those of objects with similar spectral features 
is a frequently used method to estimate the reddening. An alternative approach 
is based on the strength of narrow sodium lines in the spectrum.

\begin{figure*}
\begin{center}
   \includegraphics[width=12cm,height=7.cm]{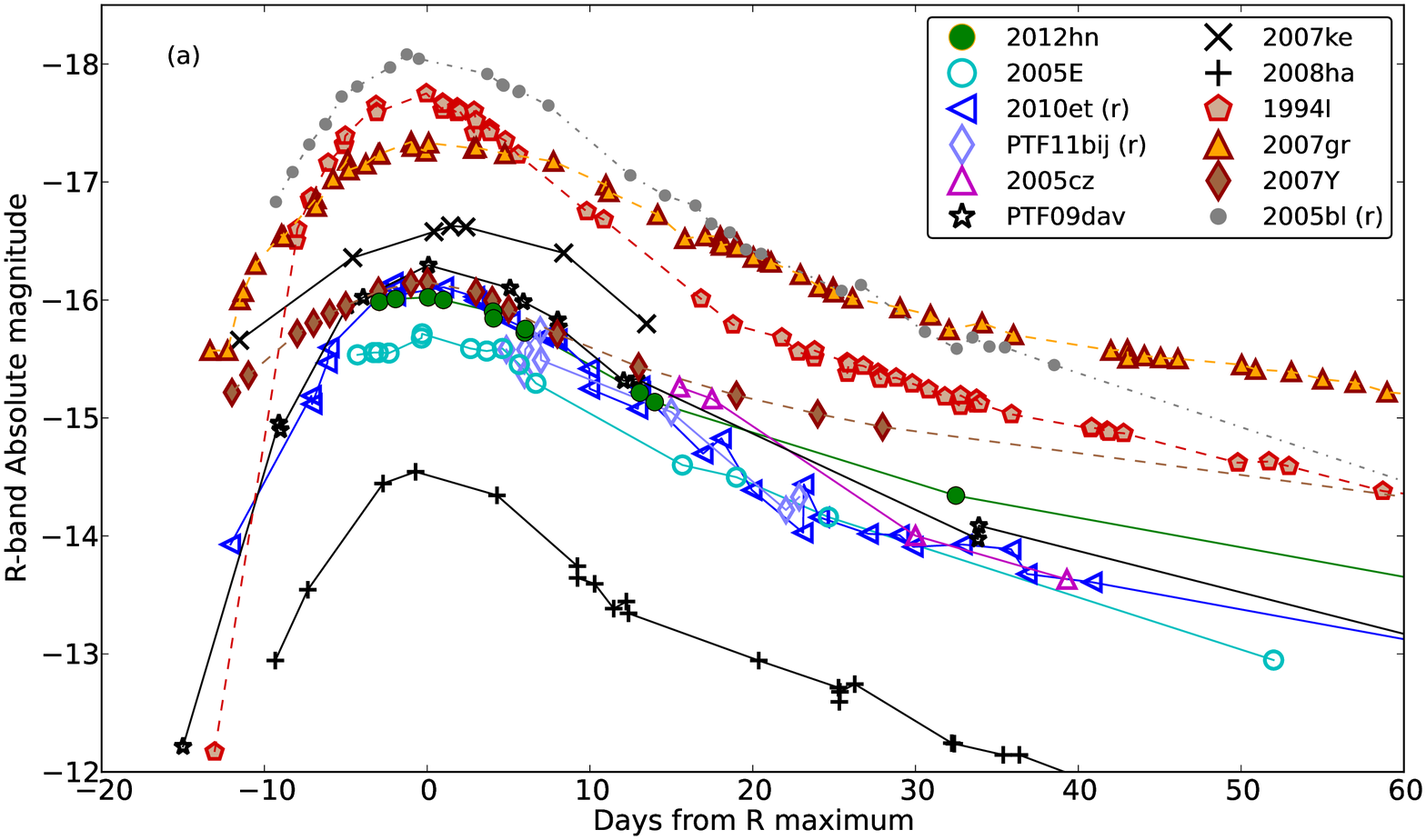}
      \includegraphics[width=12cm,height=7.cm]{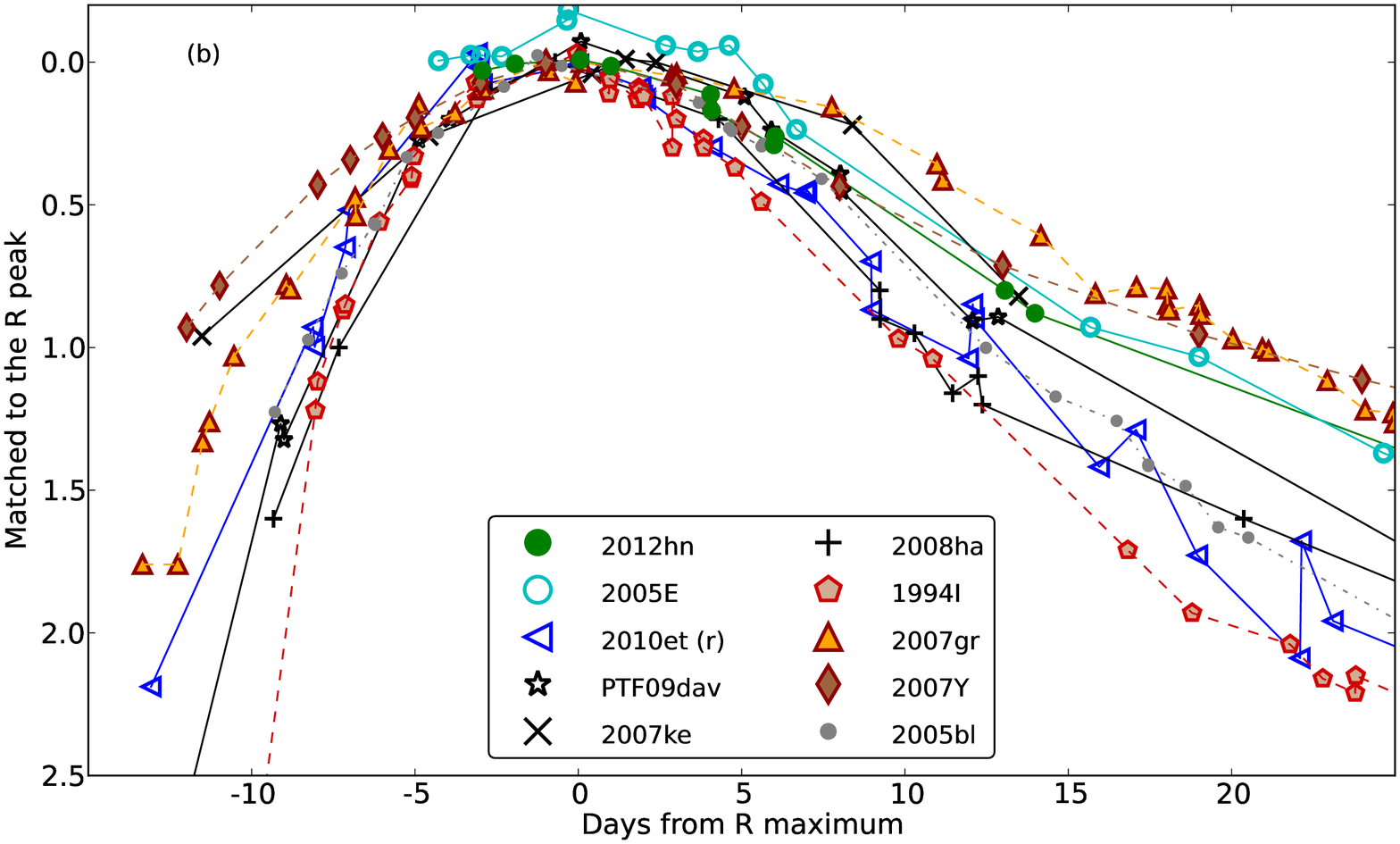}
   \caption{$R$-band photometry of SN~2012hn in comparison with other objects. 
Cool colours (open symbols) are used for faint type I SNe that share similarities 
   with SN~2012hn, black symbols for other faint type I SNe for which the link with 
   SN~2012hn is less secure, warm colours (filled symbols) for CC SNe.
   Data: SN~1994I, \protect\citet{1996AJ....111..327R};
   SN~2005E,  \protect\citet{2010Natur.465..322P}; 
   SN~2005bl, \protect\citet{2008MNRAS.385...75T}; 
   SN~2005cz, \protect\citet{2010Natur.465..326K}; 
   SN~2007Y,  \protect\citet{2009ApJ...696..713S};
   SN~2007gr, \protect\citet{2008ApJ...673L.155V} and \protect\citet{2009A&A...508..371H};       
   SN~2007ke, \protect\citet{2012ApJ...755..161K}; 
   SN~2008ha, \protect\citet{2009Natur.459..674V};
   PTF09dav,  \protect\citet{2011ApJ...732..118S};
   SN~2010et, \protect\citet{2012ApJ...755..161K}; 
   PTF11bij,  \protect\citet{2012ApJ...755..161K}.
   }
   \label{fig:lc2}
\end{center}
\end{figure*}

Gas and dust are often mixed, and some authors have claimed the existence of 
a statistical correlation between the  EW of \NaID{} lines and the colour 
excess \citep{1997A&A...318..269M,2003fthp.conf..200T}.
These results have been questioned \citep{2011MNRAS.415L..81P}, but very 
recently confirmed on the basis of an analysis of a large sample of Sloan 
spectra \citep{2012MNRAS.426.1465P}. The fact that \NaID{} is easily saturated  
makes this method inaccurate for high reddening values, while at low reddening
the spread of the relation is very high. In the XSHOOTER spectrum the \NaID{} 
doublet is identified at a redshift of 0.0076, 150 \kms{} displaced to the red  
with respect to the systemic velocity ($z=0.0071$) (Fig.~\ref{fig:naid}).  
We  measured an equivalent width of 0.55 and 1.0 \,\AA{} for the D2 and D1 
components, respectively. D1 should have twice the intensity of D2. Convolving 
our XSHOOTER spectrum with a gaussian of 15\,\AA{} (which is typical of our 
low-resolution spectra) causes the lines to disappear, which confirms that the 
non-detection in the lower-resolution spectra is not surprising.

Using the relation of \cite{2012MNRAS.426.1465P}, SN~2012hn should have an 
$E(B-V)_\mathrm{NGC\ 2272}=0.2$ mag. This reddening would make SN~2012hn 
intrinsically very similar to the faint type I SNe~2005E \citep{2010Natur.465..322P} and 
2010et \citep{2012ApJ...755..161K} in the $R$ and $I$ bands, 
while it would remain the faintest object among these three in the $B$ band 
(see Fig.~\ref{fig:lccomp}). Even though a lower reddening in the host can not be 
exluded, in this paper we adopt a host-galaxy colour excess $E(B-V)_\mathrm{NGC\ 2272}=0.2$ mag for SN~2012hn.

\section{Light curves}
\label{sec:ca}

Photometrically, SNe~2005E and 2010et provide the closest matches to SN~2012hn.
Among the other faint type I SNe, SN~2005cz \citep{2010Natur.465..326K}  
and  PTF11bij \citep{2012ApJ...755..161K} show a similar decline in the $R$ 
band, but due to poor coverage around maximum the peak luminosity is not well 
constrained (see Fig.~\ref{fig:lc2}a). PTF09dav \citep{2011ApJ...732..118S}   
has very similar luminosity in the $R$ band, while SN~2007ke 
is brighter and has a broader light curve than all the other objects of this class. 
Its connection to the faint type I transient family cannot be safely established.

It has been claimed that `Ca-rich' SNe show a faster luminosity evolution 
than other type I SNe. Their light curves likely have a rise time
$\leq$ 15 days \citep{2012ApJ...755..161K} and they seem to evolve faster 
than normal SNe Ia and most SNe~Ib/c. The rise time of SNe~Ia is $\sim$\,18 
days \citep{2011MNRAS.416.2607G,2010ApJ...712..350H}, while on average SNe~Ib 
have a rise time $\geq$ 20 days \citep{2011MNRAS.416.3138V}. However, while 
SN~2010et have a similar time evolution than SN~1994I one of the fastest 
evolving SNe~Ic ever studied, SN~2005E and SN~2012hn have a slower 
light curve  evolution that  SN 1994I and SN 2010et (see Fig. \ref{fig:lc2}b and Tab. 
\ref{tabstretch}).

\begin{table*}
  \caption{Summary of Faint type I spectro-photometric parameters.}
  \label{tabstretch}
  \footnotesize
  \begin{tabular}{@{}ccccccccc@{}}
  \hline
SN             & Peak magnitude & Stretch factor$^{a}$ & Photospheric velocity & Hydrogen & Helium & Scandium & Oxygen &  Limiting magnitude\\
             &  $R$  &   & \kms{} &  &  & &  &for a dwarf host \\
\hline 
 2012hn    &     $-15.65$ (33)  &  1.88  (20) &   10000       &   No     &  No    & No    & Yes   &   $-11$     \\
 2005E      &     $-15.68$ (10)  &  1.60  (10) &   11000       &   No     &  Yes   & No    & Yes   &   $-7.5$     \\
 2010et     &     $-16.12$ (12)  &  1.10  (10) &   10000       &   No     &  Yes   & No    & Yes   & $-12.1$    \\
 PTF11bij    &       --             &    --           &     --            &   No    &   --    & --     & Yes   & $-12.4$   \\
 2005cz     &      --              &    --           &   $\geq$11000  &  No  & Yes  & No   & Yes    & $-7.77$     \\
 2007ke    &     $-16.62$ (30) &   1.72 (20)  &   11000       &   No     &  Yes   & No    & Yes    &   --  \\
 PTF09dav  &     $-16.30$ (16) &   1.30 (19)  &   6000         &   Yes    &  No    & Yes   & No     &$-9.8$    \\
\hline
\end{tabular}\\
$^a$Stretch factor is computed using SN 1994I as reference.
\end{table*}

\begin{figure*}
\begin{center}
   \includegraphics[width=14cm,height=11cm]{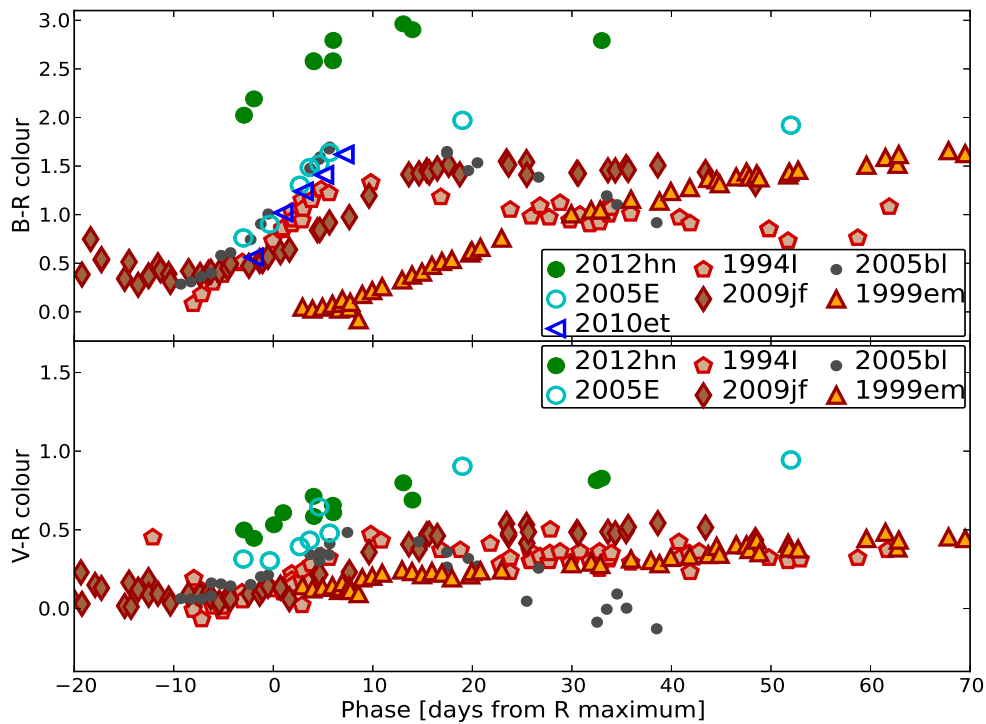}
  \caption{The $B-R$ and $V-R$ colour curves for a sample of supernovae.
  Data: SN 1999em, \protect\cite{2003MNRAS.338..939E}; SN 2005cf,
\protect\cite{2007MNRAS.376.1301P}; SN 2009jf, \protect\cite{2011MNRAS.416.3138V};
 see Fig.~\ref{fig:lc2} for the references for other SNe.}
  \label{fig:col}
\end{center}
\end{figure*}

The colour-curve comparison in Fig.~\ref{fig:col} confirms that SN~2012hn is 
very red, one magnitude redder than the closest matches, SNe~2005E and 2010et. 
All these SNe show a deficit of flux in the $B$ band and a very fast evolution 
in the blue bands, suggesting a rapid cooling of the ejecta.

\section{Spectra}
\label{sec:spe}

\begin{figure}
   \includegraphics[width=7cm,height=8.cm]{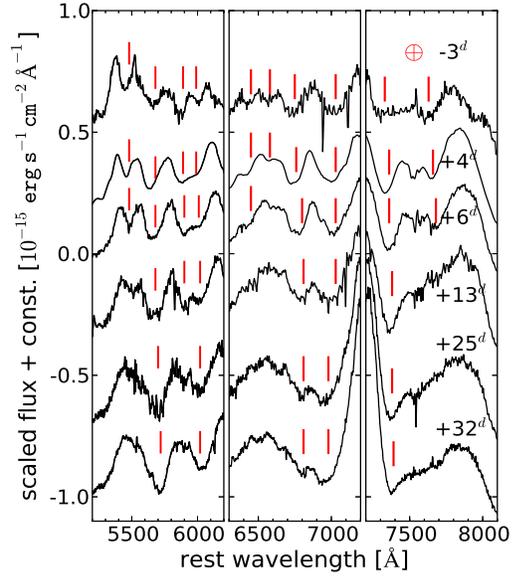}
  \caption{Detail evolution of the spectra sequence of SN~2012hn. Minima visible in the spectra have been marked at different epoch.}
  \label{fig:specinset}
\end{figure}

SN~2012hn displays significant peculiarities in its spectral evolution.  
The deficit of flux in the $B$ band is visible in all the spectra (see 
Fig.~\ref{fig:lc}). The photospheric velocity, derived from spectral fits 
with {\sc synow} (Fisher 2000), is almost constant with phase. We 
determine an expansion velocity of $\sim$\,10000 \kms{}, 
and the spectral lines appear even more blended at late times than 
at early phases (see Fig.~\ref{fig:specinset}). 
In addition, while the minima of the absorptions 
remain at the same positions, the peaks of selected emission features, 
though still blueshifted, move to redder wavelengths with time.  
In particular, the emission at $\sim$\,7195\,\AA{} (rest frame)
shifts by $\sim$\,35\,\AA{} in one month (from 7180\,\AA{} on 2012 
April 14.02  to 7215\,\AA{} on 2012 May 18.97). 
This line could be identified with forbidden [\CaII] \LL{}7291,7323. 
Its intensity increases with time (as expected for forbidden lines) 
and the blueshift goes from 5000 \kms{} in the first spectrum to 3500 
\kms{} in the last photospheric spectrum, assuming the identification 
with [\CaII] \LL{}7291,7323 is correct. Noticeably, this line is 
already visible in our first spectrum, taken before maximum light.
The unusual velocity evolution in SN 2012hn is difficult to explain. 
Normally, absorption lines in SNe tend to become narrower with 
time, as the outer layers become more diluted by the expansion 
of the ejecta and the line-forming region recedes to lower 
velocities. In SN 2012hn we observe the opposite trend: 
constant photospheric velocity (as inferred from the blueshift 
of absorption lines) and increasing line widths.

\subsection{Spectra of faint type I SNe}
\begin{figure*}
   \includegraphics[width=14.cm,height=15.cm]{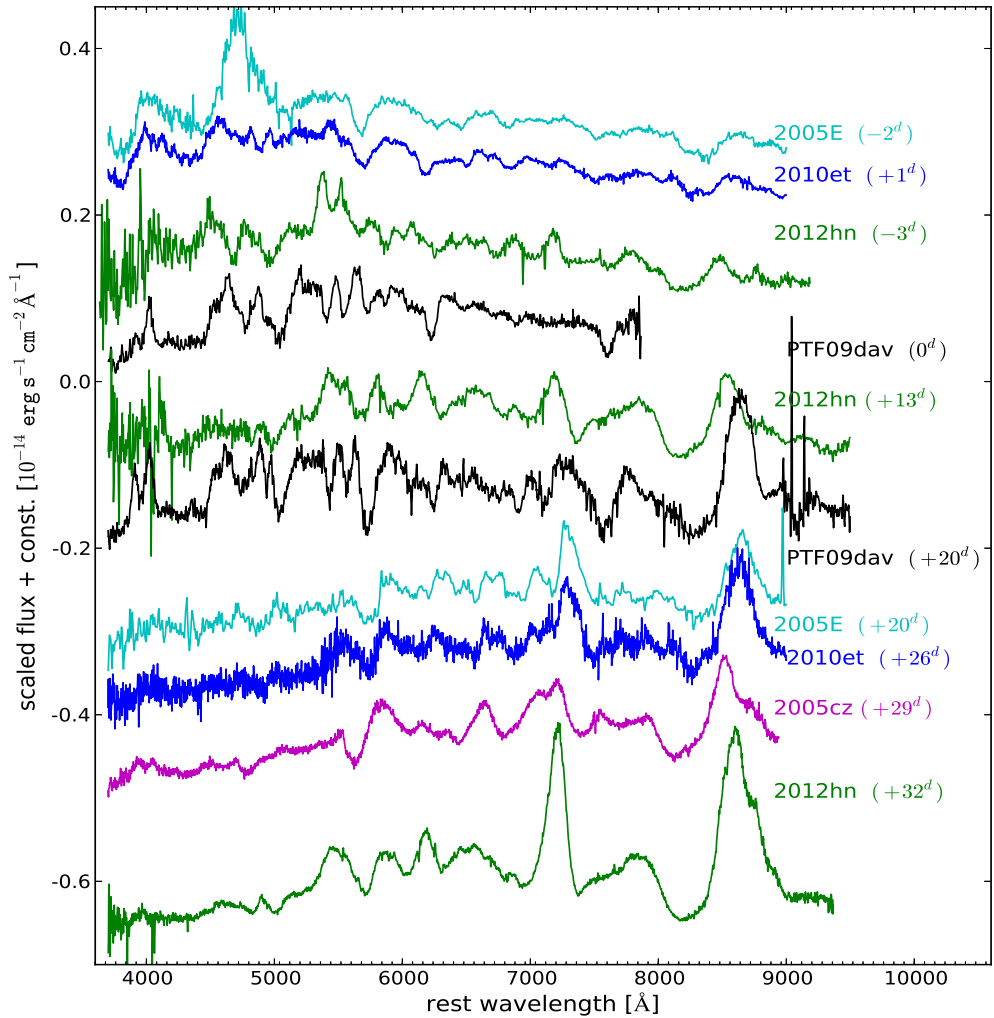}
  \caption{Spectra comparison of SNe that have been suggested as `Ca-rich'. 
  See Fig. \protect\ref{fig:lc2} for references.}
  \label{fig:speccomp1}
\end{figure*}

In Fig.~\ref{fig:speccomp1}, we show a set of spectra that includes
several faint type I objects (SNe~2005E, 2005cz and 2010et) and the 
peculiar PTF09dav. SNe~2005E and 2010et appear to be very similar, 
except for the intense peak at $\sim$\,4500\,\AA{} visible in SN~2005E.
\HeI{} is detected in both of them and strong \CaII{} lines appear 
later on. On the other hand there are several differences with the 
spectra of PTF09dav as already pointed out by \cite{2012ApJ...755..161K}.  
In particular the lack of \HeI{} features and the presence of \ScII{} 
lines make PTF09dav an outlier. We note that \HeI{} is absent also in 
the spectra of SN~2012hn, and we do not see any evidence for the presence
of \ScII{} lines. The spectra of SN~2012hn are also quite different from 
the spectra of SNe~2005E and 2010et (see Fig.~\ref{fig:speccomp1}). This 
implies a high degree of heterogeneity in the spectra of faint type I 
objects, raising the possibility that not all of these objects come 
from the same progenitor systems or share the same explosion mechanism. 

As previously mentioned, at the classification stage, we did not find 
objects in any of the publicly available archives with a spectrum similar 
to that of SN~2012hn, and we were unable to identify all the spectral 
lines. In principle, the lack of strong H and \HeI{} lines would favour 
the classification as a peculiar SN~Ic. Indeed, the comparison with a 
library of supernova spectra via {\sc gelato} \citep{2008A&A...488..383H} 
shows a best match  with SN~1990aa \citep{1990IAUC.5111....1F} and 
several other SNe~Ic.

\begin{figure}
   \includegraphics[width=8.5cm,height=7cm]{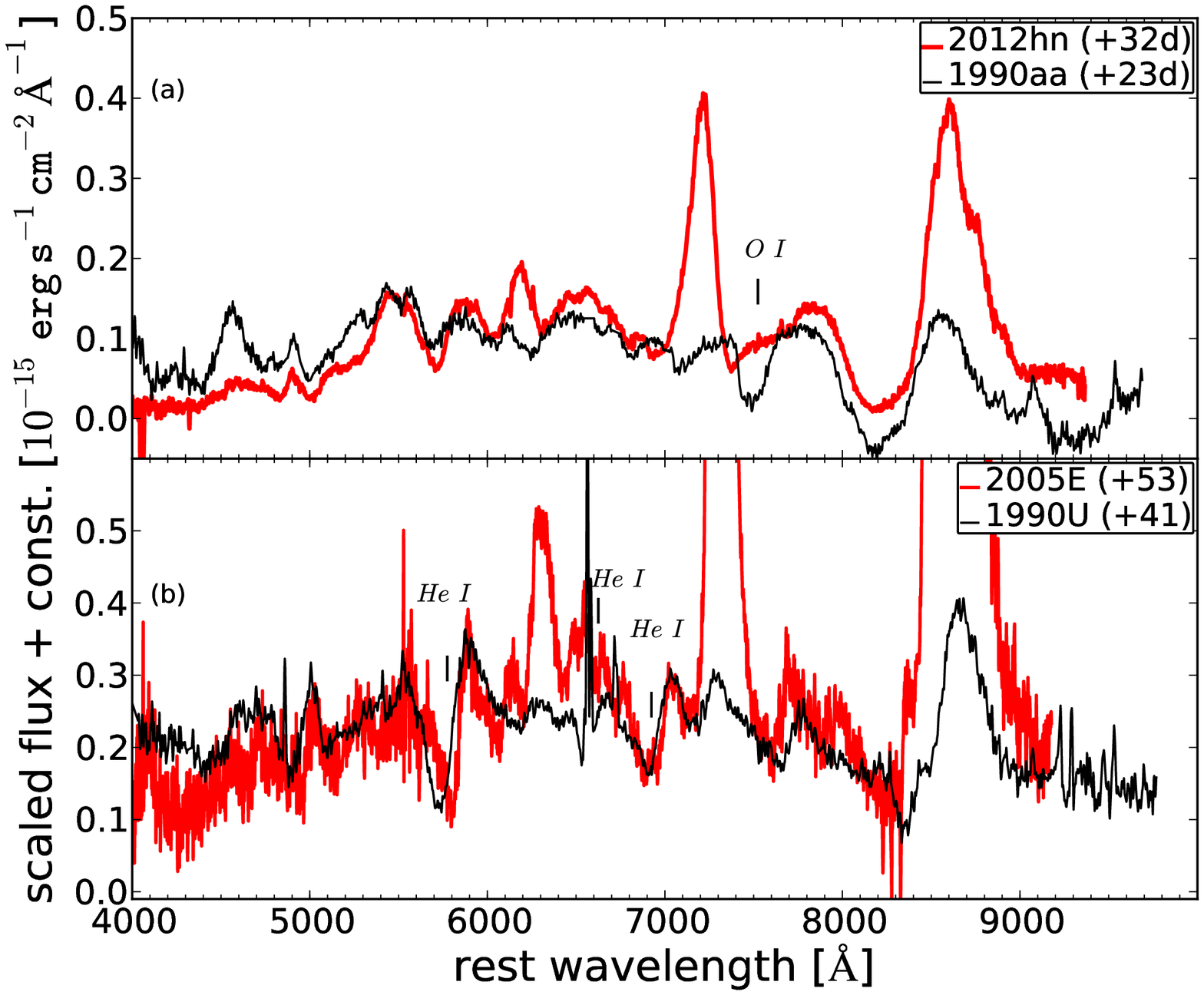}
  \caption{Comparison of SN~2012hn and SN~2005E with SN~1990aa and SN~1990U, 
  respectively. The latter have been identified as best spectroscopic matches 
  to the former. Data are from \protect\citet{2001AJ....121.1648M} and 
  \protect\citet{1992ApJ...384L..37F}. }
  \label{fig:speccomp2}
\end{figure}

In Fig.~\ref{fig:speccomp2}a we compare SN~2012hn with SN~1990aa 
\citep{2001AJ....121.1648M}.  The spectrum of SN~2012hn shows in parts 
the same features as SN~1990aa, although some emission features are much 
stronger in the former. 
This may however be due to the fact that the spectrum of SN 2012hn is more evolved.
The main differences are the strong emissions 
at $\sim$\,6100\,\AA{}, $\sim$\,7200\,\AA{} and $\sim$\,8700\,\AA{}. 
Around 7500\,\AA{} SN~1990aa shows \OI{} \L{}7774, which is not visible 
in SN~2012hn. 

In the case of SN~2005E, the most similar object was identified
in SN~1990U \citep{2010Natur.465..322P}. A comparison is shown
in Fig.~\ref{fig:speccomp2}b. 
Both SN~2005E and SN~1990U  show \HeI{} \L{}5876, \L{}7065 lines.
 In the spectrum of SN~1990U  \HeI{} \L{}6678 is detected, although contaminated by
narrow  \Ha{} from the host galaxy,
while it is weak or absent in the spectrum of SN~2005E.
  
It is interesting to point out that both SN~2012hn and SN~2005E are 
similar to quite normal SNe Ic and SNe Ib, respectively, though with 
more prominent Ca features. 
Given these similarities, it is tempting to propose that faint type I 
SNe might arise from similar progenitors as SNe Ib/c, though with an 
overproduction of Ca in the explosive nucleosynthesis or with 
ionisation and excitation conditions in the ejecta that favour 
the formation of strong Ca II features.

\begin{figure}
  \includegraphics[width=8.5cm,height=8.5cm]{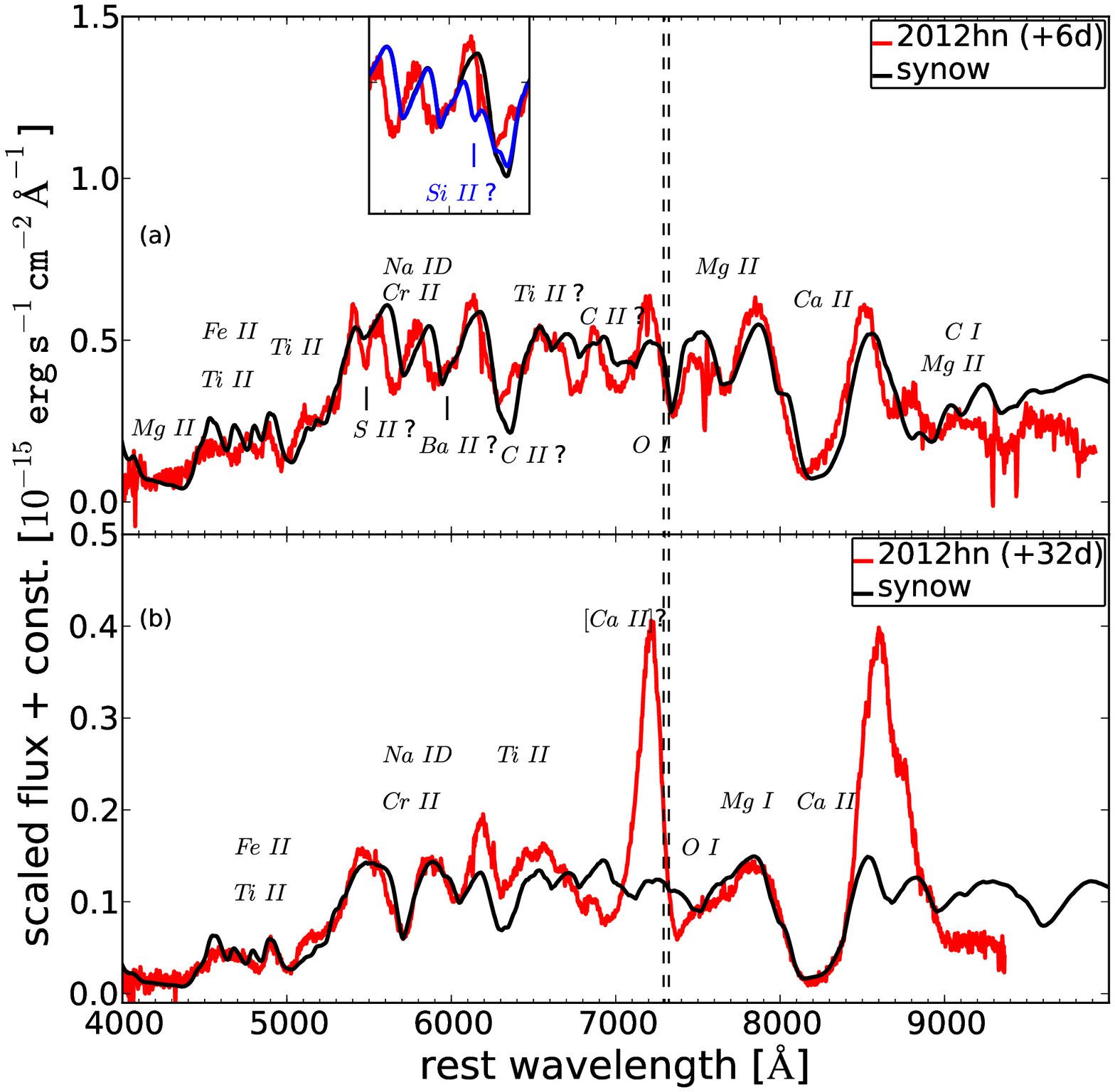}
  \caption{\protect{\sc synow} fit of SN~2012hn spectra. The dashed lines 
  indicate the rest-frame positions of [\CaII] \protect\LL{}7291,7323. 
  The inset in the upper panel shows that adding \SiII{} (blue line)
  deteriorates the  \protect{\sc synow} fit to the observed spectrum.}
  \label{fig:speccomp3}
\end{figure}

Starting from the similarity with  SN~1990aa, we used {\sc synow}
to identify  the lines  in the SN~2012hn spectra. Fits to the spectra 
collected on April 20th (+4 days) and May 18th (+32 days) are shown  
in Figs.~\ref{fig:speccomp3}a and \ref{fig:speccomp3}b, respectively. 
A photospheric velocity of 10000 \kms{} was adopted in both cases, 
while we used temperatures of 4500 K for the spectrum at +6 days 
and 4200 K for the spectrum at +32 days.
The radial dependence of the line optical depths was chosen to be 
exponential with the $e$-folding velocity $v_e$ [i.e. $\tau \propto 
\exp(-v/v_e)$] set to 1000 \kms{} for both spectra. 
The spectra can be reproduced reasonably well with a small number of 
ions, including \FeII{}, \CaII{}, \CI{}, \CrII{}, \TiII{} and \NaI{}. 
Most lines between 6000 and 7500\,\AA{} can be reproduced by enhancing 
\FeII{} and \TiII{}. The line at $\sim$\,5800\,\AA{} may be reproduced 
by \NaI{} and \CrII{} or detached \HeI{}, but we consider the \HeI{} 
identification unlikely because of the lack of other \HeI{} lines both 
in the optical and the NIR regime (see Sec. \ref{nir}). Some \FeII{} lines may
also contribute to this feature.  
Between 7200 and 7800\,\AA{} SN~2012hn shows two  
absorption features, redward and blueward respectively of the expected 
position of \OI{} \L{}7774.  \MgII{} is an alternative for the 
absorption at 7700\,\AA{}, redward of the usual \OI{} position. The 
absorption at $\sim$\,7350\,\AA{}, blueward of the usual \OI{} position 
may still be explained with \OI{}. We stress that \SiII{} is not required to reproduce 
the spectrum (see inset of Fig. \ref{fig:speccomp3}a), while some \SII{} might possibly contribute 
to the absorption at $\sim$ 5500\,\AA{}.
In the first spectrum including some \BaII{} may help 
to reproduce the feature at 5900\,\AA{}. The presence of \BaII{},
if confirmed, would be important, since \BaII{} is not expected in 
thermonuclear explosions. Unfortunately, we consider this identification
not as secure as other identifications, since it is mainly based on a single line.

The main shortcomings of the synthetic spectrum are the emission 
features at $\sim$\,7200\,\AA{} and $\sim$\,8700\,\AA{} which are not 
reproduced. If the former ($\sim$\,7200\,\AA{}) is identified as  
blueshifted [\CaII] or [\OII], the poor fit is not surprising,  
since {\sc synow} can not reproduce forbidden lines.

\subsection{SN~2012hn NIR spectrum}
\label{nir}

\begin{figure}
   \includegraphics[width=8.5cm,height=7cm]{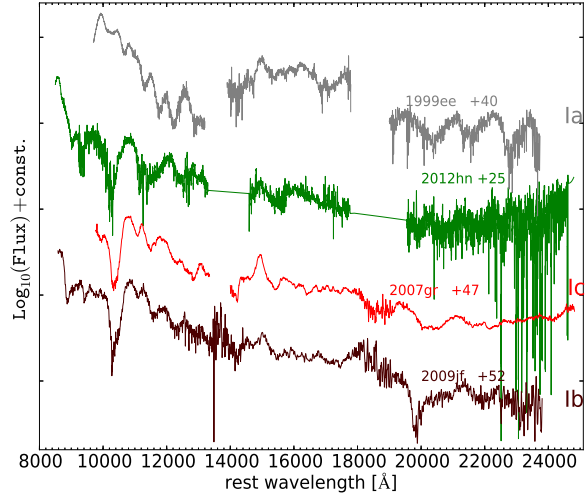}
  \caption{XSHOOTER NIR spectrum of SN~2012hn in comparison with spectra 
  of other SNe. The SN~2012hn spectrum has been rebinned to 20\,\AA{} bins. 
  Comparison spectra are from \protect\cite{2002AJ....124..417H,2009A&A...508..371H,2011MNRAS.413.2583S}.} 
  \label{fig:specinf}
\end{figure}

The XSHOOTER spectrum  of SN~2012hn collected on 2012 May 11.98 for 
the first time gives us the possibility to probe the NIR spectrum
of a faint type I SN (see Fig.~\ref{fig:specinf}). The red part of 
the spectrum is quite noisy, but still it can be used to exclude the
presence of a strong He feature at $\sim$\,2 micron, which is typical 
of He-rich SNe. Normal SNe~Ia show intense lines of \CoII{} and \FeII{} 
at this phase \citep{2009AJ....138..727M,2012arXiv1208.5949G}.
These lines are not clearly identified in SN~2012hn. However, this is not
unexpected, since the low luminosity of SN~2012hn suggests that a
 small amount of $^{56}$Ni was synthesized in the 
explosion. The XSHOOTER spectrum also gives us information on the 
amount of flux emitted in the NIR. After checking the XSHOOTER spectrum 
flux calibration with our optical photometry, we computed a synthetic 
$R-H$ colour ($\sim$\,$-0.5$ to 0.0 mag). This value is comparable with 
the $R-H$ colour of several SNe Ia \citep{2004AJ....128.3034K}, 
and much lower than the $R-H$ colour of core-collapse SNe at this phase 
($R-H$\,$\sim$\,1 mag).

\subsection{Late-time spectroscopy}
\label{sec:latespectrum}

\begin{figure*}
  \includegraphics[width=14cm,height=15cm]{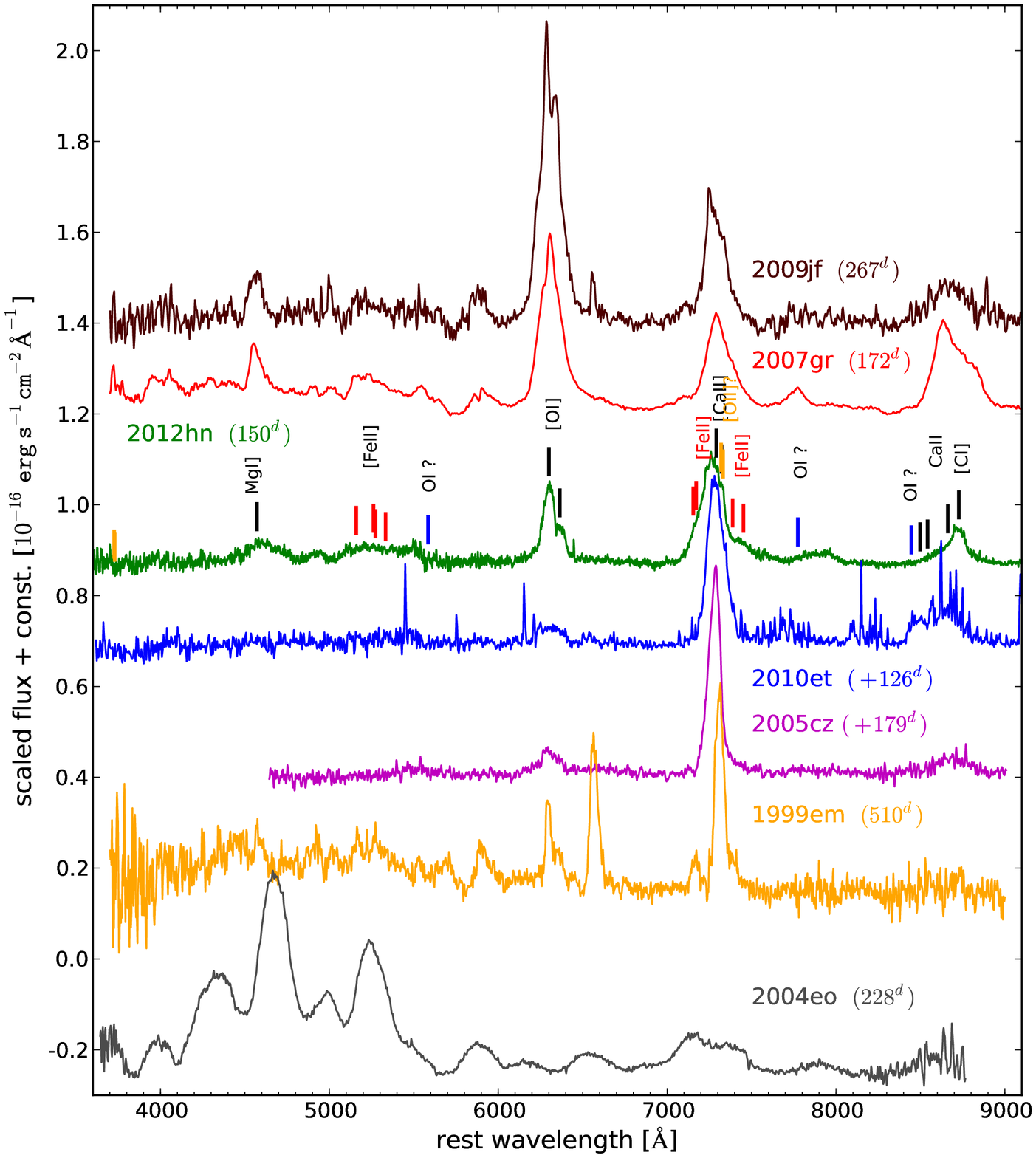}
  \caption{Nebular spectrum of SN~2012hn in comparison with a set of nebular 
  spectra of different types of SNe. Tentative line identifications are given 
  for the spectrum of SN~2012hn. Data: SN~2009jf, \protect\citet{2011MNRAS.416.3138V}; 
  SN~2007gr, \protect\citet{2009A&A...508..371H}; SN~2010et, \protect\citet{2012ApJ...755..161K}; 
  SN~2005cz, \protect\citet{2010Natur.465..326K}; SN~1999em, \protect\citet{2003MNRAS.338..939E}; 
  SN~2004eo, \protect\cite{2007MNRAS.377.1531P}.}
  \label{fig:latespec}
\end{figure*}

Two XSHOOTER spectra were obtained 149 and 150 days after $R$-band maximum, 
and the combination of their optical parts is compared with nebular spectra 
of different types of SNe in Fig.~\ref{fig:latespec}. No flux has been 
detected in the NIR part of the combined XSHOOTER spectrum down to a mag 
of $\sim$\,22.5 in the $H$ band. 

The emission centred at $\sim$\,7250\,\AA{} is most likely [\CaII] 
\LL{}7291,7323, with a contribution of [\FeII] to the blue (\LL{}7155,7172) 
and the red (\LL{}7388,7452) wings. \MgI] \L{}4571 is also detected. 
No permitted lines of \OI{} (\L{}7774 or \L{}8446) are visible, and also the 
Ca NIR triplet, very strong at early phases, has faded. An emission feature 
at $\sim$\,8700\,\AA{} is likely a  blend of residual, weak \CaII{} and 
[\CI] \L{}8727. The absence of permitted lines gives us a constraint on the 
density, confirming that the spectrum is nebular \citep{1989ApJ...343..323F}.  
Comparing SN~2012hn with other faint type I SNe, the weakness of the \CaII{} 
NIR triplet is not unusual. However, the  prominent [\OI] \LL{}6300,6364 
feature is quite unique among faint type I SNe. It is much more similar in 
terms of relative strength to spectra of stripped core-collapse SNe,
which show strong [\OI] owing to oxygen produced in the progenitor star.
The feature is symmetric around the rest wavelength and relatively 
narrow, with emission out to ~3500 \kms. 
No narrow lines from the host galaxy are visible in the spectrum.
Using eq. 2 from \cite{Kennicutt1998} and an \Ha{} upper limit of 3 x $10^{-18}$
\ene{} cm$^{-2}$, we obtained an upper limit for the star formation rate of 2 x $10^{-6}$ 
\msun{} yr$^{-1}$

\begin{figure}
  \includegraphics[width=9cm,height=7.cm]{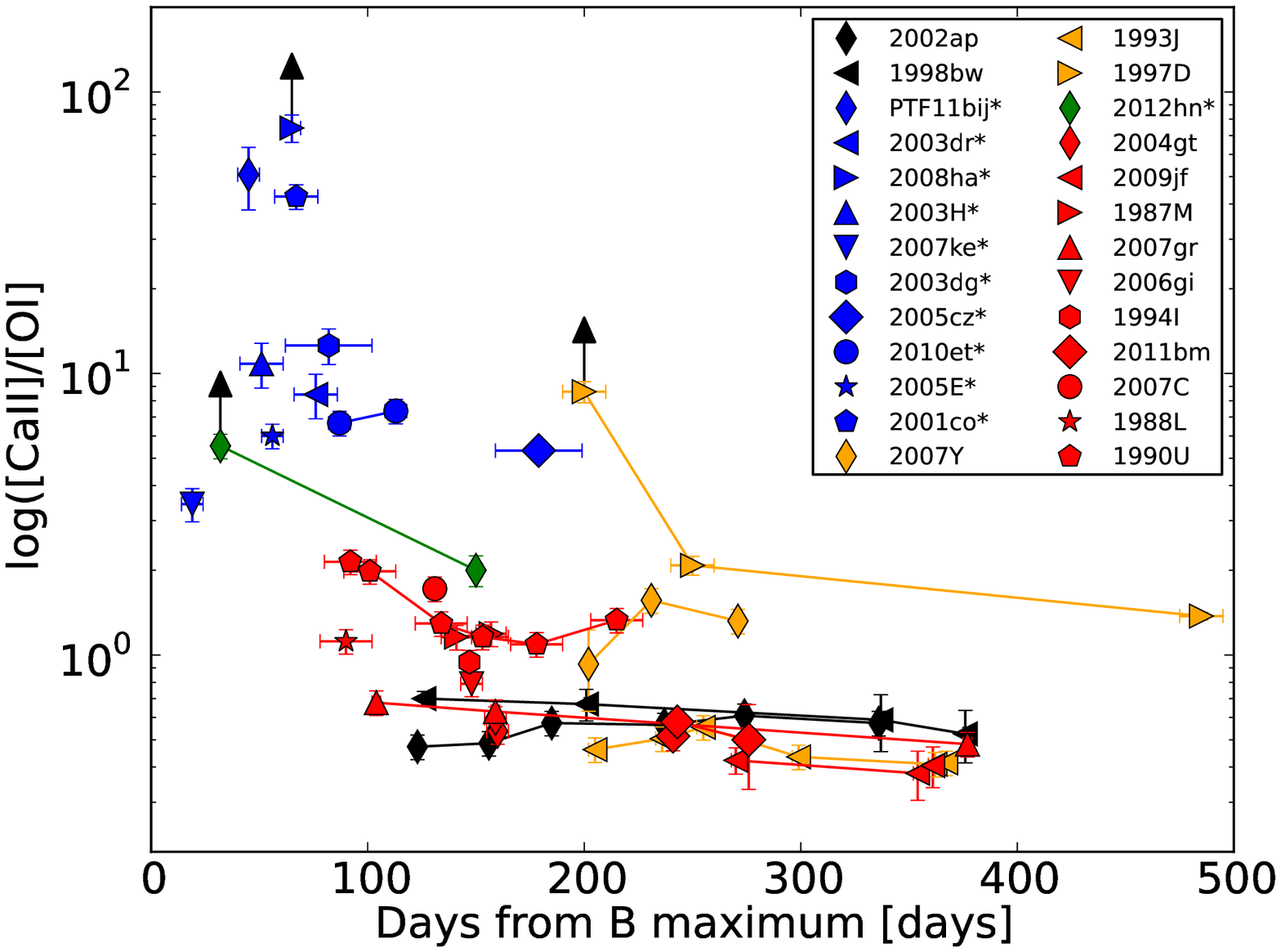}
  \caption{The [\CaII{}]/[\OI{}] ratio as a function of time for a sample of SNe:
  stripped-envelope CC SNe (in red), broad-line stripped-envelope CC SNe (in black), 
  He/H-rich CC SNe (in yellow), SN~2012hn (in green) and other faint type I SNe 
  (in blue). For objects labelled with $*$ the phase is given relative to $R$-band maximum 
  or relative to discovery. Data from \protect\citet[][and references therein]{2009MNRAS.397..677T}, 
  \protect\citet{2009Natur.459..674V}, \protect\citet{2009A&A...508..371H},
   \protect\cite{2009ApJ...696..713S}, \protect\citet{2011MNRAS.416.3138V}, 
 \protect\citet{2012ApJ...749L..28V},  \protect\citet{2012ApJ...755..161K}, \protect\citet{1998ApJ...498L.129T}.}
  \label{fig:CaO}
\end{figure}

We computed the [\CaII{}]/[\OI{}] ratio for a large set of nebular spectra 
of core-collapse SNe and the sample of faint type I SNe presented by 
\cite{2012ApJ...755..161K} (see Fig.~\ref{fig:CaO}). For almost all 
core-collapse SNe [\CaII] \LL{}7291,7323 starts to be visible earlier than 
[\OI] \LL{}6300,6364, but rarely less than 100 days after maximum light. 
Faint type I SNe show both these lines earlier on, but no data are 
available later than 170 days after maximum to study the intensity evolution 
of these features over longer time scales. Fig.~\ref{fig:CaO} confirms that 
at 150 days after maximum the [\CaII{}]/[\OI{}] ratio of 
SN~2012hn is at the edge of the region occupied by core-collapse SNe.

\section{Bolometric light curve}
\label{sec:bolometric}

To constrain the physical parameters of SN~2012hn ($M_\mathrm{Ni}$, 
$M_\mathrm{ej}$ and $E_\mathrm{k}$), we computed a pseudo-bolometric light 
curve using the available photometric information. As long as the light 
curve is powered by the $^{56}$Ni $\rightarrow ^{56}$Co $\rightarrow ^{56}$Fe 
decay chain, a brighter bolometric light curve implies a larger amount 
of ejected $^{56}$Ni. In addition, the broader the light curve, the 
higher the ejected mass and/or the lower the kinetic energy released in 
the explosion \citep{1982ApJ...253..785A}.

\begin{figure*}
   \includegraphics[width=16cm,height=12cm]{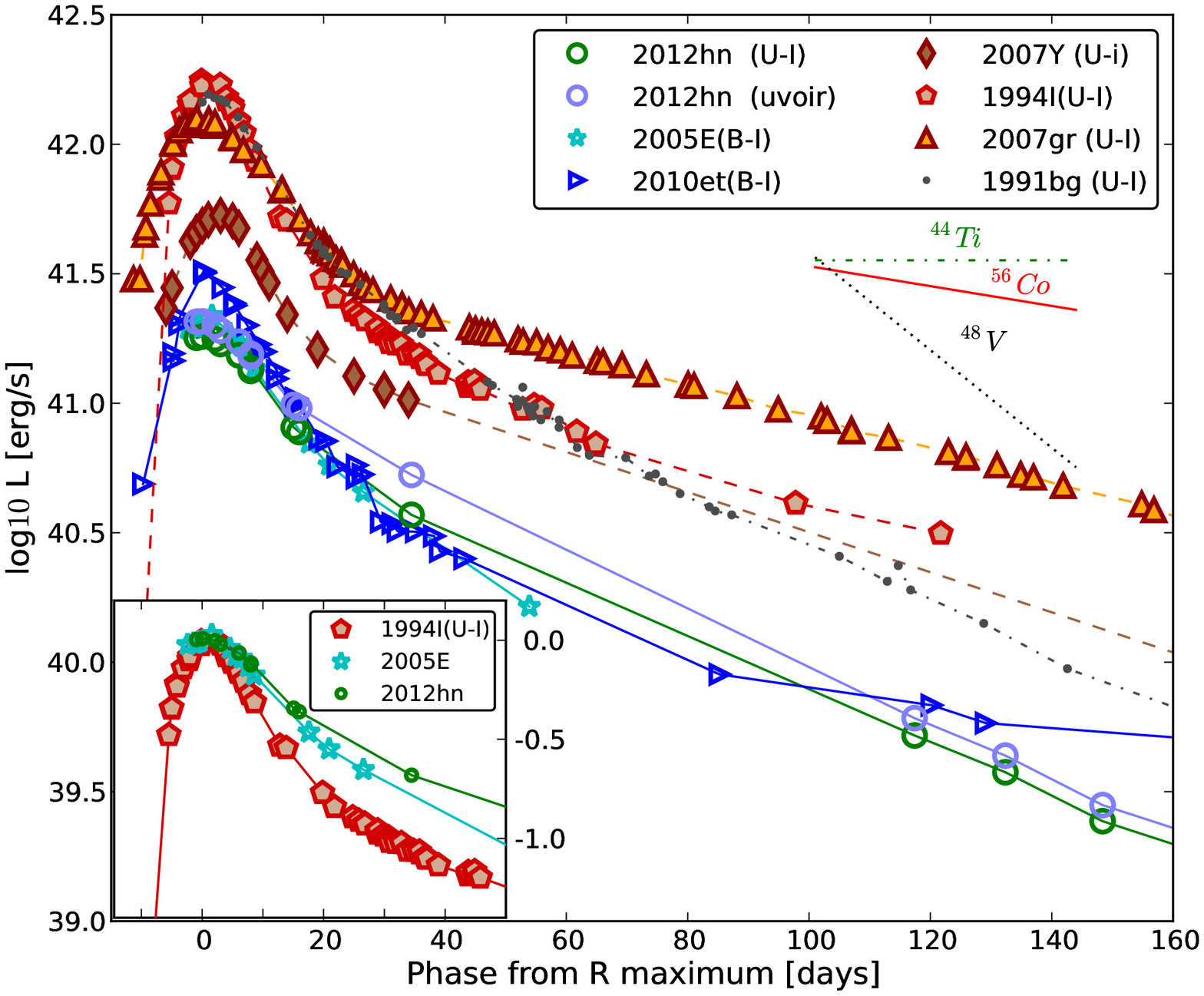}
  \caption{Pseudo-Bolometric light curve of SN~2012hn in comparison with other SNe.
  Data: see references in Fig.~\protect\ref{fig:lc2}; SN1991bg, \protect\cite{1996MNRAS.283....1T}.
  Distances and colour excesses:
  - SN 1994I:  $E(B-V) =$ 0.04 mag,  $\mu=$ 29.60 mag, \protect\citep{2006MNRAS.369.1939S};
  - SN~2007gr: $E(B-V) =$ 0.092 mag, $\mu=$ 29.84 mag, \protect\citep{2008ApJ...673L.155V};
  - SN~2007Y:  $E(B-V) =$ 0.112 mag, $\mu=$ 31.13 mag, \protect\citep{2009ApJ...696..713S};
  - SN~2005E:  $E(B-V) =$ 0.041 mag, $\mu=$ 32.66 mag, \protect\citep{2010Natur.465..322P};
  - SN~2010et: $E(B-V) =$ 0.046 mag, $\mu=$ 35.05 mag, \protect\citep{2012ApJ...755..161K};
  - SN 1991bg: $E(B-V) =$ 0.029 mag, $\mu=$ 31.44 mag, \protect\citep{1996MNRAS.283....1T}. Inset plot: pseudo-bolometric light curves of SNe 2012hn, 2005E and 1994 normalized to the maximum luminosity.}
  \label{fig:bolom}
\end{figure*}

To compute the pseudo-bolometric light curve, the observed 
magnitudes were corrected for reddening ($E(B-V)_\mathrm{NGC\ 2272}=0.2$ mag), 
converted to flux densities at the effective wavelengths
and integrated using Simpson’s rule.
The so-created pseudo-bolometric light curve is 
plotted in Fig.~\ref{fig:bolom} together with those of other SNe computed 
following the same prescriptions. In core collapse SNe the NIR flux 
contributes up to 40--50\,$\%$ to the total flux, in SNe~Ia up to 30\,$\%$ 
\citep{2008MNRAS.383.1485V}. Given the $R-H$ colour measured from our 
first XSHOOTER spectrum, we assumed that the fractional contributions of 
the UV and NIR emission to the bolometric light curve of SN~2012hn are 
the same as for the type Ia SN 2005cf \citep{2007MNRAS.376.1301P}
With this assumption, we computed a $uvoir$ pseudo-bolometric light curve 
of SN~2012hn, which is also shown in Fig.~\ref{fig:bolom}. 

The lack of data before maximum and hence the unknown rise time of 
SN~2012hn makes it difficult to derive accurate explosion parameters.
A rough estimate of $M_\mathrm{ej}$ and $E_\mathrm{k}$ can be computed 
guided by the comparison with well-studied SNe and the following simple 
relations:  $v \propto (E_\mathrm{k}/M_\mathrm{ej})^{1/2}$  and  
$t_\mathrm{s} \propto (M^{3}_\mathrm{ej}/E_\mathrm{k})^{1/4}$,
where $v$ is the photospheric velocity (evaluated at maximum light) 
and $t_\mathrm{s}$ is the time scale 
of the photospheric phase \citep{1982ApJ...253..785A}.  
Using a Chandrasekhar-mass SN~Ia as comparison ($E_\mathrm{k}=1.3 
\times 10^{51}$ erg, $M_\mathrm{ej}=1.4$ \msun{}, $v=10000$ \kms{}, the photospheric 
velocity of SN~2012hn ($v=10000$ \kms{}) and a time scale of 
$t_\mathrm{s}(\mathrm{SN~2012hn})=$ 0.7--0.8 $t_\mathrm{s}(\mathrm{SNe~Ia})$ 
(from light-curve comparisons), we obtain the following values for 
SN~2012hn: $M_\mathrm{ej}=$ 0.7--0.9 \msun{} 
and $E_\mathrm{k}=$ 0.65--0.85 $\times 10^{51}$ erg. 
Using instead the type Ic SN~2007gr as reference, adopting the values of 
$M_\mathrm{ej}$ and  $E_\mathrm{k}$ reported by \cite{2009A&A...508..371H},  
we obtain the following estimate for the energy and 
the ejected mass: $M_\mathrm{ej}=$ 1.2--2.1 \msun{} and $E_\mathrm{k}=$ 
0.74--3 $\times 10^{51}$ erg. 

\citet{2010Natur.465..322P}, using the same approach, obtained much 
smaller values for SN~2005E ($M_\mathrm{ej}=$ 0.25--0.41 \msun{} 
and $E_\mathrm{k}= $0.44--0.72 $\times 10^{51}$ erg). 
Some of the difference may come from the presence of helium in SN 2005E.
As reported by \cite{2010Natur.465..322P}, helium is less opaque than other 
elements, and this may lead to an underestimate of the helium mass.
But the different values obtained mainly reflect the uncertainties in some of 
the important fitting constraints, such as the rise time. \cite{2010Natur.465..322P} indeed 
adopted a very short rise time of 7--9 days.

We suggest that from the light-curve comparison with SN~1994I 
\citep[with a rise time of 12 days, ][]{Iwamoto1994}; 
see inset plot in Fig.~\ref{fig:bolom}), a rise 
time of 7--9 days for SN~2005E is probably an underestimate, unless the light curve of 
SN~2005E was strongly asymmetric. Using a larger rise time for SN~2005E would give 
ejected mass and kinetic energy larger than previously reported, more consistent with 
those we obtain for SN~2012hn.
However, a spectral model for a not completely 
nebular spectrum of SN~2005E was presented by \protect\citet{2010Natur.465..322P},
supporting a low ejected mass and kinetic energy for SN~2005E.
Future observations of faint type I SNe are needed to confirm or disprove 
the very small ejected masses and kinetic energies proposed for some
faint type I SNe.

>From the absolute luminosity of SN~2012hn, if the light curve is powered 
by nickel decay, assuming a rise time between 10 and 15 days,
the mass of $^{56}$Ni produced in the explosion should be in the range 
$M_\mathrm{Ni}$ = 0.005-0.010 \msun{}, where  0.005 \msun{} is the 
lower limit considering no infrared contribution to the bolometric light curve. 
Assuming that the fractional contribution of the UV and NIR emission to the 
bolometric light curve is 40--50$\%$, a value typical for core collapse SNe, 
we can fix an upper limit for the nickel produced in the exposion of  0.013 \msun{}.

However, as pointed out by several authors, in low-density explosions 
decays of radioactive nuclei other than $^{56}$Ni may be important  
\citep{2010ApJ...715..767S,2011ApJ...738...21W,2012MNRAS.420.3003S}. In 
particular the decays ${}^{48}$Cr $\rightarrow ^{48}$V $\rightarrow ^{48}$Ti 
and ${}^{44}$Ti $\rightarrow ^{44}$Sc $\rightarrow ^{44}$Ca may play an important 
role in powering the light curve. 
While at early phases it is tricky to identify the radioactive nuclei 
that contribute to power the light curve, the slope of the light curve  at later phases
provides more information.
In the former case, the slope of the light curve would be comparable 
or even steeper (in case of incomplete trapping of the energy) to the slope of the 
 $^{48}$V decay (half life time = 15.97 days), while in the latter case, the ${}^{44}$Ti decay 
 (half life time = 63 years) would determine the observed light curve slope. 
 From the inferred slope of the bolometric light curve of SN~2012hn, 
 we consider a major contribution from these alternative radioactive nuclei 
 to the light curve unlikely.
  
The nickel mass produced may hence be even lower than the one reported. 
Ti and Cr may also be responsible for the fast evolution in the blue 
bands and the relatively red colour by depressing most of the flux in 
the blue part of the spectrum. The slope of the bolometric light curve 
in the tail phase is comparable to that of the type Ia SN~1991bg and 
faster than in most other SNe. SN~2010et, the only `Ca-rich' SN with 
multi-band late-time photometry, shows a similar slope until 80 days 
after $R$-band maximum. After this point, the bolometric light curve 
of SN~2010et is contaminated by the flux from its host galaxy 
\citep{2012ApJ...755..161K}. The steady slope of the tail of SN~2012hn 
suggests that a single decay is powering the light curve at these phases.

\section{Discussion and Conclusions}
\label{sec:discussion}
SN~2012hn belongs to the class of faint type I SNe, showing a light curve 
that is very similar to those of SNe~2005E and 2010et. It has a low peak 
luminosity and evolves rapidly (although not as fast as SN~1994I). Faint 
type I transients show heterogeneous spectral properties. In particular, 
SN~2012hn shows no clear evidence of \HeI{} features, and the early-time 
spectra resemble those of a SN~Ic with superimposed forbidden lines of 
\CaII{}. Contrary to what we typically see in SNe, line widths tend to 
increase with time, while the photospheric velocity remains almost constant 
at $\sim$\,10000 \kms{} during the entire photospheric phase.

One of the most intriguing features is the presence, very early on, of 
a strong emission feature at about 7200 \AA{}. Based on the very strong 
\CaII{} NIR triplet and the lack of viable alternatives, we attribute 
this feature to forbidden \CaII{}. However, if this is correct, the peak 
of the emission is  blueshifted by 3000--5000 \kms{} with respect to the 
rest-frame position. Since the blueshift decreases with time, it is unlikely
due to dust forming  and may be explained by an optically thick core
of the ejecta \citep{2009MNRAS.397..677T} or by strong asymmetry 
in the explosion.
On the other hand, the [\OI] profile in the nebular spectrum is consistent 
with a spherically symmetric distribution of the ejecta, or at 
least of the oxygen-rich  material.

In the last few years several studies have been carried out to explain 
faint type I transients. Most of them have focused on He-shell detonations 
on accreting carbon-oxygen white dwarfs (WDs) \citep{2007ApJ...662L..95B,
2010ApJ...715..767S,2011ApJ...738...21W,2012MNRAS.420.3003S}.

In the double-detonation scenario for SNe~Ia, the He-shell detonation 
is followed by a second detonation in the core of the carbon-oxygen WD. 
While it is still not clear if and where the second detonation occurs 
\citep[but see ][]{2007A&A...476.1133F,2010A&A...514A..53F},
a pure He-shell detonation might  produce an explosion with several 
characteristics similar to the explosion of SN~2012hn.
\citet{2010ApJ...715..767S} investigated He-shell detonations for three 
different WD masses (0.6, 1.0, 1.2 \msun) and different He-shell masses
(0.02, 0.05, 0.1, 0.2, 0.3 \msun). \citet{2011ApJ...738...21W} focused 
their study on low-mass WDs (0.45--0.6 \msun) with a 0.2 \msun{} He shell. 
Also \citet{2012MNRAS.420.3003S} studied He-shell detonations on low-mass 
WDs (0.45-0.58 \msun) with a 0.21 \msun{} He shell, but extended the 
simulations to cases where a second detonation occurs. 
These studies tend to find that 
the amount of  intermediate-mass elements (Si, S) produced during 
the explosion  is lower than in \emph{normal} SNe Ia whereas Ti  and Cr
are produced abundantly.
The resultant spectrum is also rich of  Ti and Cr lines.
While this is consistent with the spectra of  SN~2012hn, 
the light curve may be different. If the second detonation occurs,
the models show light curves brighter than that of SN~2012hn
\citep{2012MNRAS.420.3003S}. If the second detonation  does not occur,
the peak luminosity is comparable, but the evolution is much faster 
than that of SN~2012hn.
\cite{2011ApJ...738...21W} were able to obtain a light curve comparable
to SN~2005E for their He-shell detonation model 
WD (0.45 \msun) + He-shell (0.2 \msun), but only 
with an artificial $^{44}$Ti enhancement by a factor of 50.
All the He-shell detonation models in low-mass 
WDs predict red spectra with Ti lines and no intermediate-mass elements,
but none of them predict the slow spectral evolution of SN~2012hn. 
So far, no synthetic nebular spectra of He detonations on 
accreting WDs have been made.
Unburned C/O can be present in a He-shell detonation if the second
detonation in the WD does not occur.
The He-shell detonation model of 
\cite{2012MNRAS.420.3003S} shows oxygen below 4000 \kms{},
consistent with the oxygen velocity in the nebular spectraum of SN~2012hn.
Whether this amount of oxygen is enough and the density and ionisation in the 
innermost ejecta are suited to produce the observed nebular oxygen line of 
SN~2012hn is still not clear.
An estimate of the ejected mass of oxygen and carbon from the nebular SN~2012hn spectrum would 
be the next step in this analysis \citep{2007ApJ...670..592M,2010MNRAS.408...87M}.

Little Si and S, the presence of Ti and Cr, and [OI] in the 
nebular spectrum are also expected in core-collapse SNe. Nevertheless, only few
theoretical studies have been made so far to explain faint type I SNe as 
core-collapse explosions. 
\citet{2010Natur.465..326K} proposed for SN~2005cz 
a progenitor of 8--12 \msun{} that lost its envelope through interaction in 
a binary system. These stars are more abundant than more massive stars, and 
some of them should still be present in E/S0 galaxies. If the hydrogen layer 
has been stripped by a companion star, they might still produce core-collapse 
SNe similar to SNe~Ib and Ic. As stripped-envelope counterparts of faint type IIP SNe 
\citep{2004MNRAS.347...74P}, a population of faint SNe Ib/c may be expected 
to exist. These objects have not been unambiguously identified in observations 
so far \citep[but see ][]{2009Natur.459..674V}, but faint type I SNe could be 
viable candidates. The weak point of this interpretation is that, so far, no faint type IIP SNe have been discovered in remote locations of E/S0 galaxies. 
Only few models for faint stripped-envelope core-collapse SNe are available. \cite{2009ApJ...707..193F} presented first radiation-hydrodynamics calculations of several models for faint SNe including fallback SNe and accretion-induced collapse (AIC) SNe. Their conclusion was that both AIC and failed thermonuclear SNe can potentially reproduce the observed light curves of faint type I SNe, but that further studies are recommended.

In summary, the physical origin of this faint type I SN~2012hn is still 
ambiguous. The location of the transient, in the remote outskirts of an 
E/S0 type galaxy indicates an origin in an old stellar population and hence 
an accreting WD system as the most likely progenitor. The lack of any 
possible dwarf host galaxy to quite faint magnitudes in prediscovery images 
(to below about $M = -11$) argues either for a non-starforming population 
or an ultra-faint dwarf galaxy with presumably low metallicity. However, 
the nebular specturm of SN~2012hn at +150d shows the strongest oxygen 
feature yet detected among faint type I SNe, similar in strength to [\OI] 
seen in core-collapse SNe. 
This is not necessary surprising, since the nebular spectrum of SN 2012hn 
is among most evolved and it is well known that the  [\OI] feature  becomes 
more intense with time. 
This could also mean that the high [\CaII]/[\OI] ratio seen in some faint type I SNe 
at relative early phase may in part be a phase-dependent effect rather than 
reflecting an intrinsic abundance pattern.
\MgI{} and \CI{} features are also detected in SN~2012hn, which are weak or 
absent in other faint type I SNe, but usually seen in core-collapse events. 

Among the other faint type I SNe discovered so far, SNe~2010et and 2005E 
have similar light curves and are in even more remote locations with respect 
to their hosts. This appears to be a strong argument in favour of the WD as 
progenitors. However, if a single scenario gives rise to these events, it 
must be able to reproduce the differences in their spectra, in particular 
the small amount of helium (if any) in SN~2012hn and the differences 
already mentioned in the nebular spectra, and to produce an ejected mass 
in the range ($\sim$\,0.5--1.5 \msun), comparable with the ejected masses 
of low-mass, low-energy SNe Ib/c. Detailed modelling of the nebular spectra 
of this class may help to further constrain the origins of these peculiar 
and intriguing explosions 
\citep[e.g.][]{2010MNRAS.408...87M,2011MNRAS.416..881M,2011MNRAS.418.1517M,2012A&A...546A..28J}
and to confirm or disprove the large Ca abundance reported for SN~2005E by \citet{2010Natur.465..322P}.

Including other faint transients, such as PTF09dav or SN~2008ha, in the 
same scenario will be even more complicated, emphasizing the possibility 
that faint type I SNe may actually arise from several different explosion 
channels.

\section*{Acknowledgements}
G.P. acknowledges support by the Proyecto FONDECYT 11090421.
J. A. acknowledges support by CONICYT through FONDECYT grant 3110142. 
G.P. and J.A. acknowledge support by the Millennium Center for Supernova 
Science (P10-064-F), with input from Fondo de Innovacion para la 
Competitividad, del Ministerio de Economia, Fomento y Turismo de Chile.
S.T. acknowledges support by the TRR 33 `The Dark Universe' of the German Research Foundation.
MS acknowledges support from the Royal Society.
S.G. acknowledges support from FONDECYT through grant 3130680.
This work was supported by the National Science Foundation under 
grants PHY 11-25915 and AST 11-09174. 
Research leading to these results has received funding from the 
European Research Council under the European Union's Seventh 
Framework Programme (FP7/2007-2013)/ERC Grant agreement 
n$^{\rm o}$ [291222]  (PI : S. J. Smartt). A.P. S.B. and E.C. are partially 
supported by the PRIN-INAF 2011 with the project "Transient 
Universe: from ESO Large to PESSTO". Research by AGY and his group 
is supported by the FP7/ERC, Minerva and GIF grants.
Parts of this research were conducted by the Australian Research 
Council Centre of Excellence for All-sky Astrophysics (CAASTRO), 
through project number CE110001020. We are grateful to Douglas 
Leonard who provided us his spectrum of SN~2005cz.  

This paper is based on observations made with the following facilities: 
ESO Telescopes at the La  Silla and Paranal Observatories under programme IDs 
184.D-1140, 188.D-3003, 089.D-0270,
Prompt telescopes (Chile). Trappist telescope (Chile), Du Pont telescope 
(Chile), Magellan telescope (Chile).
Observations under programme ID 188.D-3003 are part of PESSTO 
(the Public ESO Spectroscopic Survey for Transient Objects).
The spectra of \cite{2012ApJ...755..161K} were downloaded from 
WISeREP \protect\citep{2012PASP..124..668Y}. 
  

\begin{thebibliography}{76}
\expandafter\ifx\csname natexlab\endcsname\relax\def\natexlab#1{#1}\fi

\bibitem[{Anderson \& James(2009)}]{2009MNRAS.399..559A}
Anderson J.~P., James P.~A., 2009, Monthly Notices of the Royal Astronomical
  Society, 399, 559

\bibitem[{Arnett(1982)}]{1982ApJ...253..785A}
Arnett W.~D., 1982, The Astrophysical Journal, 253, 785

\bibitem[{Benitez-Herrera {et~al}\mbox{.}(2012)Benitez-Herrera, Taubenberger,
  Valenti, Benetti, \& Pastorello}]{2012ATel.4047....1B}
Benitez-Herrera S., Taubenberger S., Valenti S., Benetti S., Pastorello A.,
  2012, The Astronomer's Telegram, 4047, 1

\bibitem[{Bildsten {et~al}\mbox{.}(2007)Bildsten, Shen, Weinberg, \&
  Nelemans}]{2007ApJ...662L..95B}
Bildsten L., Shen K.~J., Weinberg N.~N., Nelemans G., 2007, The Astrophysical
  Journal, 662, L95

\bibitem[{Drake {et~al}\mbox{.}(2010)Drake, Djorgovski, Mahabal, Grahamm,
  Williams, Prieto, Catelan, Beshore, Larson, \&
  Christensen}]{2010CBET.2339....1D}
Drake A.~J. {et~al.}, 2010, Central Bureau Electronic Telegrams, 2339, 1

\bibitem[{Elmhamdi {et~al}\mbox{.}(2003)Elmhamdi, Danziger, Chugai, Pastorello,
  Turatto, Cappellaro, Altavilla, Benetti, Patat, \&
  Salvo}]{2003MNRAS.338..939E}
Elmhamdi A. {et~al.}, 2003, Monthly Notice of the Royal Astronomical Society,
  338, 939

\bibitem[{Filippenko(1992)}]{1992ApJ...384L..37F}
Filippenko A.~V., 1992, The Astrophysical Journal, 384, L37

\bibitem[{Filippenko {et~al}\mbox{.}(2003)Filippenko, Chornock, Swift, Modjaz,
  Simcoe, \& Rauch}]{2003IAUC.8159....2F}
Filippenko A.~V., Chornock R., Swift B., Modjaz M., Simcoe R., Rauch M., 2003,
  IAU Circ., 8159, 2

\bibitem[{Filippenko, Shields \& Petschek(1990)Filippenko, Shields, \&
  Petschek}]{1990IAUC.5111....1F}
Filippenko A.~V., Shields J.~C., Petschek A.~G., 1990, IAU Circ., 5111, 1

\bibitem[{Filippenko {et~al}\mbox{.}(2007)Filippenko, Silverman, Foley, Modjaz,
  Papovich, Willmer, Blondin, \& Brown}]{2007CBET.1101....1F}
Filippenko A.~V., Silverman J.~M., Foley R.~J., Modjaz M., Papovich C., Willmer
  C. N.~A., Blondin S., Brown W., 2007, Central Bureau Electronic Telegrams,
  1101, 1

\bibitem[{Fink, Hillebrandt \& R\"{o}pke(2007)Fink, Hillebrandt, \&
  R\"{o}pke}]{2007A&A...476.1133F}
Fink M., Hillebrandt W., R\"{o}pke F.~K., 2007, Astronomy and Astrophysics,
  476, 1133

\bibitem[{Foley {et~al}\mbox{.}(2013)Foley, Challis, Chornock, Ganeshalingam,
  Li, Marion, Morrell, Pignata, Stritzinger, Silverman, Wang, Anderson,
  Filippenko, Freedman, Hamuy, Jha, Kirshner, McCully, Persson, Phillips,
  Reichart, \& Soderberg}]{Foley2013}
Foley R.~J. {et~al.}, 2013, The Astrophysical Journal, 767, 57

\bibitem[{Foley {et~al}\mbox{.}(2009)Foley, Chornock, Filippenko,
  Ganeshalingam, Kirshner, Li, Cenko, Challis, Friedman, Modjaz, Silverman, \&
  Wood-Vasey}]{2009AJ....138..376F}
Foley R.~J. {et~al.}, 2009, The Astronomical Journal, 138, 376

\bibitem[{F\"{o}rster \& Schawinski(2008)}]{2008MNRAS.388L..74F}
F\"{o}rster F., Schawinski K., 2008, Monthly Notices of the Royal Astronomical
  Society: Letters, 388, L74

\bibitem[{Fransson \& Chevalier(1989)}]{1989ApJ...343..323F}
Fransson C., Chevalier R.~A., 1989, The Astrophysical Journal, 343, 323

\bibitem[{Fryer {et~al}\mbox{.}(2009)Fryer, Brown, Bufano, Dahl, Fontes, Frey,
  Holland, Hungerford, Immler, Mazzali, Milne, Scannapieco, Weinberg, \&
  Young}]{2009ApJ...707..193F}
Fryer C.~L. {et~al.}, 2009, The Astrophysical Journal, 707, 193

\bibitem[{Gall {et~al}\mbox{.}(2012)Gall, Taubenberger, Kromer, Sim, Benetti,
  Blanc, Elias-Rosa, \& Hillebrandt}]{2012arXiv1208.5949G}
Gall E. E.~E., Taubenberger S., Kromer M., Sim S.~A., Benetti S., Blanc G.,
  Elias-Rosa N., Hillebrandt W., 2012, arXiv.org, 1208, 5949

\bibitem[{Ganeshalingam, Li \& Filippenko(2011)Ganeshalingam, Li, \&
  Filippenko}]{2011MNRAS.416.2607G}
Ganeshalingam M., Li W., Filippenko A.~V., 2011, Monthly Notices of the Royal
  Astronomical Society, 416, 2607

\bibitem[{Hamuy {et~al}\mbox{.}(2002)Hamuy, Maza, Pinto, Phillips, Suntzeff,
  Blum, Olsen, Pinfield, Ivanov, Augusteijn, Brillant, Chadid, Cuby, Doublier,
  Hainaut, {Le Floc'h}, Lidman, Petr-Gotzens, Pompei, \&
  Vanzi}]{2002AJ....124..417H}
Hamuy M. {et~al.}, 2002, The Astronomical Journal, 124, 417

\bibitem[{Harutyunyan {et~al}\mbox{.}(2008)Harutyunyan, Pfahler, Pastorello,
  Taubenberger, Turatto, Cappellaro, Benetti, Elias-Rosa, Navasardyan, Valenti,
  Stanishev, Patat, Riello, Pignata, \& Hillebrandt}]{2008A&A...488..383H}
Harutyunyan A.~H. {et~al.}, 2008, Astronomy and Astrophysics, 488, 383

\bibitem[{Hayden {et~al}\mbox{.}(2010)Hayden, Garnavich, Kessler, Frieman, Jha,
  Bassett, Cinabro, Dilday, Kasen, Marriner, Nichol, Riess, Sako, Schneider,
  Smith, \& Sollerman}]{2010ApJ...712..350H}
Hayden B.~T. {et~al.}, 2010, The Astrophysical Journal, 712, 350

\bibitem[{Hunter {et~al}\mbox{.}(2009)Hunter, Valenti, Kotak, Meikle,
  Taubenberger, Pastorello, Benetti, Stanishev, Smartt, Trundle, Arkharov,
  Bufano, Cappellaro, {Di Carlo}, Dolci, Elias-Rosa, Frandsen, Fynbo, Hopp,
  Larionov, Laursen, Mazzali, Navasardyan, Ries, Riffeser, Rizzi, Tsvetkov,
  Turatto, \& Wilke}]{2009A&A...508..371H}
Hunter D.~J. {et~al.}, 2009, Astronomy and Astrophysics, 508, 371

\bibitem[{Iwamoto {et~al}\mbox{.}(1994)Iwamoto, Nomoto, Hoflich, Yamaoka,
  Kumagai, \& Shigeyama}]{Iwamoto1994}
Iwamoto K., Nomoto K., Hoflich P., Yamaoka H., Kumagai S., Shigeyama T., 1994,
  The Astrophysical Journal, 437, L115

\bibitem[{Jerkstrand {et~al}\mbox{.}(2012)Jerkstrand, Fransson, Maguire,
  Smartt, Ergon, \& Spyromilio}]{2012A&A...546A..28J}
Jerkstrand A., Fransson C., Maguire K., Smartt S., Ergon M., Spyromilio J.,
  2012, Astronomy and Astrophysics, 546, 28

\bibitem[{Kaiser {et~al}\mbox{.}(2002)Kaiser, Aussel, Burke, Boesgaard,
  Chambers, Chun, Heasley, Hodapp, Hunt, Jedicke, Jewitt, Kudritzki, Luppino,
  Maberry, Magnier, Monet, Onaka, Pickles, Rhoads, Simon, Szalay, Szapudi,
  Tholen, Tonry, Waterson, \& Wick}]{Kaiser2002}
Kaiser N. {et~al.}, 2002, in Survey and Other Telescope Technologies and
  Discoveries. Edited by Tyson, Tyson J.~A., Wolff S., eds., Vol. 4836, pp.
  154--164

\bibitem[{Kasliwal {et~al}\mbox{.}(2010)Kasliwal, Kulkarni, Gal-Yam, Yaron,
  Quimby, Ofek, Nugent, Poznanski, Jacobsen, Sternberg, Arcavi, Howell,
  Sullivan, Rich, Burke, Brimacombe, Milisavljevic, Fesen, Bildsten, Shen,
  Cenko, Bloom, Hsiao, Law, Gehrels, Immler, Dekany, Rahmer, Hale, Smith,
  Zolkower, Velur, Walters, Henning, Bui, \& McKenna}]{2010ApJ...723L..98K}
Kasliwal M.~M. {et~al.}, 2010, The Astrophysical Journal Letters, 723, L98

\bibitem[{Kasliwal {et~al}\mbox{.}(2012)Kasliwal, Kulkarni, Gal-Yam, Nugent,
  Sullivan, Bildsten, Yaron, Perets, Arcavi, Ben-Ami, Bhalerao, Bloom, Cenko,
  Filippenko, Frail, Ganeshalingam, Horesh, Howell, Law, Leonard, Li, Ofek,
  Polishook, Poznanski, Quimby, Silverman, Sternberg, \&
  Xu}]{2012ApJ...755..161K}
Kasliwal M.~M. {et~al.}, 2012, The Astrophysical Journal, 755, 161

\bibitem[{Kawabata {et~al}\mbox{.}(2010)Kawabata, Maeda, Nomoto, Taubenberger,
  Tanaka, Deng, Pian, Hattori, \& Itagaki}]{2010Natur.465..326K}
Kawabata K.~S. {et~al.}, 2010, Nature, 465, 326

\bibitem[{Kelly, Kirshner \& Pahre(2008)Kelly, Kirshner, \&
  Pahre}]{2008ApJ...687.1201K}
Kelly P.~L., Kirshner R.~P., Pahre M., 2008, The Astrophysical Journal, 687,
  1201

\bibitem[{Kennicutt(1998)}]{Kennicutt1998}
Kennicutt R.~C., 1998, The Astrophysical Journal, 498, 541

\bibitem[{Krisciunas {et~al}\mbox{.}(2004)Krisciunas, Suntzeff, Phillips,
  Candia, Prieto, Antezana, Chassagne, Chen, Dickinson, Eisenhardt, Espinoza,
  Garnavich, Gonz\'{a}lez, Harrison, Hamuy, Ivanov, Krzeminski, Kulesa,
  McCarthy, Moro-Mart\'{\i}n, Muena, Noriega-Crespo, Persson, Pinto, Roth,
  Rubenstein, Stanford, Stringfellow, Zapata, Porter, \&
  Wischnjewsky}]{2004AJ....128.3034K}
Krisciunas K. {et~al.}, 2004, The Astronomical Journal, 128, 3034

\bibitem[{Kromer {et~al}\mbox{.}(2010)Kromer, Sim, Fink, R\"{o}pke, Seitenzahl,
  \& Hillebrandt}]{2010A&A...514A..53F}
Kromer M., Sim S.~a., Fink M., R\"{o}pke F.~K., Seitenzahl I.~R., Hillebrandt
  W., 2010, Astronomy and Astrophysics, 514, A53

\bibitem[{Leloudas {et~al}\mbox{.}(2011)Leloudas, Gallazzi, Sollerman,
  Stritzinger, Fynbo, Hjorth, Malesani, Micha$\backslash$lowski,
  Milvang-Jensen, \& Smith}]{2011A&A...530A..95L}
Leloudas G. {et~al.}, 2011, Astronomy and Astrophysics, 530, 95

\bibitem[{Lyman {et~al}\mbox{.}(2013)Lyman, James, Perets, Anderson, Gal-Yam,
  Mazzali, \& Percival}]{Lyman2013}
Lyman J.~D., James P.~A., Perets H.~B., Anderson J.~P., Gal-Yam A., Mazzali P.,
  Percival S.~M., 2013, Monthly Notices of the Royal Astronomical Society, -1,
  16

\bibitem[{Marion {et~al}\mbox{.}(2009)Marion, H\"{o}flich, Gerardy, Vacca,
  Wheeler, \& Robinson}]{2009AJ....138..727M}
Marion G.~H., H\"{o}flich P., Gerardy C.~L., Vacca W.~D., Wheeler J.~C.,
  Robinson E.~L., 2009, The Astronomical Journal, 138, 727

\bibitem[{Matheson {et~al}\mbox{.}(2001)Matheson, Filippenko, Li, Leonard, \&
  Shields}]{2001AJ....121.1648M}
Matheson T., Filippenko A.~V., Li W., Leonard D.~C., Shields J.~C., 2001, The
  Astronomical Journal, 121, 1648

\bibitem[{Maurer {et~al}\mbox{.}(2011)Maurer, Jerkstrand, Mazzali,
  Taubenberger, Hachinger, Kromer, Sim, \& Hillebrandt}]{2011MNRAS.418.1517M}
Maurer I., Jerkstrand A., Mazzali P.~A., Taubenberger S., Hachinger S., Kromer
  M., Sim S., Hillebrandt W., 2011, Monthly Notices of the Royal Astronomical
  Society, 418, 1517

\bibitem[{Mazzali {et~al}\mbox{.}(2007)Mazzali, Kawabata, Maeda, Foley, Nomoto,
  Deng, Suzuki, Iye, Kashikawa, Ohyama, Filippenko, Qiu, \&
  Wei}]{2007ApJ...670..592M}
Mazzali P.~A. {et~al.}, 2007, The Astrophysical Journal, 670, 592

\bibitem[{Mazzali {et~al}\mbox{.}(2011)Mazzali, Maurer, Stritzinger,
  Taubenberger, Benetti, \& Hachinger}]{2011MNRAS.416..881M}
Mazzali P.~A., Maurer I., Stritzinger M., Taubenberger S., Benetti S.,
  Hachinger S., 2011, Monthly Notices of the Royal Astronomical Society, 416,
  881

\bibitem[{Mazzali {et~al}\mbox{.}(2010)Mazzali, Maurer, Valenti, Kotak, \&
  Hunter}]{2010MNRAS.408...87M}
Mazzali P.~A., Maurer I., Valenti S., Kotak R., Hunter D., 2010, Monthly
  Notices of the Royal Astronomical Society, 408, 87

\bibitem[{Metzger {et~al}\mbox{.}(2009)Metzger, Piro, Quataert, \&
  Thompson}]{2009arXiv0908.1127M}
Metzger B.~D., Piro A.~L., Quataert E., Thompson T.~A., 2009, arXiv.org, 0908,
  6

\bibitem[{Moriya {et~al}\mbox{.}(2010)Moriya, Tominaga, Tanaka, Nomoto, Sauer,
  Mazzali, Maeda, \& Suzuki}]{2010ApJ...719.1445M}
Moriya T., Tominaga N., Tanaka M., Nomoto K., Sauer D.~N., Mazzali P.~A., Maeda
  K., Suzuki T., 2010, The Astrophysical Journal, 719, 1445

\bibitem[{Munari \& Zwitter(1997)}]{1997A&A...318..269M}
Munari U., Zwitter T., 1997, Astronomy and Astrophysics

\bibitem[{Nomoto, Thielemann \& Yokoi(1984)Nomoto, Thielemann, \&
  Yokoi}]{Nomoto1984}
Nomoto K., Thielemann F.-K., Yokoi K., 1984, The Astrophysical Journal, 286,
  644

\bibitem[{Pastorello {et~al}\mbox{.}(2007{\natexlab{a}})Pastorello, Mazzali,
  Pignata, Benetti, Cappellaro, Filippenko, Li, Meikle, Arkharov, Blanc,
  Bufano, Derekas, Dolci, Elias-Rosa, Foley, Ganeshalingam, Harutyunyan, Kiss,
  Kotak, Larionov, Lucey, Napoleone, Navasardyan, Patat, Rich, Ryder, Salvo,
  Schmidt, Stanishev, Sz\'{e}kely, Taubenberger, Temporin, Turatto, \&
  Hillebrandt}]{2007MNRAS.377.1531P}
Pastorello A. {et~al.}, 2007{\natexlab{a}}, Monthly Notices of the Royal
  Astronomical Society, 377, 1531

\bibitem[{Pastorello {et~al}\mbox{.}(2007{\natexlab{b}})Pastorello,
  Taubenberger, Elias-Rosa, Mazzali, Pignata, Cappellaro, Garavini, Nobili,
  Anupama, Bayliss, Benetti, Bufano, Chakradhari, Kotak, Goobar, Navasardyan,
  Patat, Sahu, Salvo, Schmidt, Stanishev, Turatto, \&
  Hillebrandt}]{2007MNRAS.376.1301P}
Pastorello A. {et~al.}, 2007{\natexlab{b}}, Monthly Notices of the Royal Astronomical Society,
  376, 1301

\bibitem[{Pastorello {et~al}\mbox{.}(2004)Pastorello, Zampieri, Turatto,
  Cappellaro, Meikle, Benetti, Branch, Baron, Patat, Armstrong, Altavilla,
  Salvo, \& Riello}]{2004MNRAS.347...74P}
Pastorello A. {et~al.}, 2004, Monthly Notices of the Royal Astronomical Society, 347, 74

\bibitem[{Perets {et~al}\mbox{.}(2010)Perets, Gal-Yam, Mazzali, Arnett, Kagan,
  Filippenko, Li, Arcavi, Cenko, Fox, Leonard, Moon, Sand, Soderberg, Anderson,
  James, Foley, Ganeshalingam, Ofek, Bildsten, Nelemans, Shen, Weinberg,
  Metzger, Piro, Quataert, Kiewe, \& Poznanski}]{2010Natur.465..322P}
Perets H.~B. {et~al.}, 2010, Nature, 465, 322

\bibitem[{Poznanski {et~al}\mbox{.}(2010)Poznanski, Chornock, Nugent, Bloom,
  Filippenko, Ganeshalingam, Leonard, Li, \& Thomas}]{2010Sci...327...58P}
Poznanski D. {et~al.}, 2010, Science (New York, N.Y.), 327, 58

\bibitem[{Poznanski {et~al}\mbox{.}(2011)Poznanski, Ganeshalingam, Silverman,
  \& Filippenko}]{2011MNRAS.415L..81P}
Poznanski D., Ganeshalingam M., Silverman J.~M., Filippenko A.~V., 2011,
  Monthly Notices of the Royal Astronomical Society: Letters, 415, L81

\bibitem[{Poznanski, Prochaska \& Bloom(2012)Poznanski, Prochaska, \&
  Bloom}]{2012MNRAS.426.1465P}
Poznanski D., Prochaska J.~X., Bloom J.~S., 2012, Monthly Notices of the Royal
  Astronomical Society, 426, 1465

\bibitem[{Pumo {et~al}\mbox{.}(2009)Pumo, Turatto, Botticella, Pastorello,
  Valenti, Zampieri, Benetti, Cappellaro, \& Patat}]{2009ApJ...705L.138P}
Pumo M.~L. {et~al.}, 2009, The Astrophysical Journal Letters, 705, L138

\bibitem[{{Quimby}(2006)}]{QuimbyRobertMichael2006}
{Quimby, Robert Michael}, 2006, ProQuest Dissertations And Theses; Thesis
  (Ph.D.)--The University of Texas at Austin

\bibitem[{Rau {et~al}\mbox{.}(2009)Rau, Kulkarni, Law, Bloom, Ciardi,
  Djorgovski, Fox, Gal-Yam, Grillmair, Kasliwal, Nugent, Ofek, Quimby, Reach,
  Shara, Bildsten, Cenko, Drake, Filippenko, Helfand, Helou, Howell, Poznanski,
  \& Sullivan}]{2009PASP..121.1334R}
Rau A. {et~al.}, 2009, Publications of the Astronomical Society of the Pacific,
  121, 1334

\bibitem[{Reichart {et~al}\mbox{.}(2005)Reichart, Nysewander, Moran, Bartelme,
  Bayliss, Foster, Clemens, Price, Evans, Salmonson, Trammell, Carney, Keohane,
  \& Gotwals}]{2005NCimC..28..767R}
Reichart D. {et~al.}, 2005, Il Nuovo Cimento C, 28, 767

\bibitem[{Richmond {et~al}\mbox{.}(1996)Richmond, van Dyk, Ho, Peng, Paik,
  Treffers, Filippenko, Bustamante-Donas, Moeller, Pawellek, Tartara, \&
  Spence}]{1996AJ....111..327R}
Richmond M.~W. {et~al.}, 1996, Astronomical Journal v.111, 111, 327

\bibitem[{Sahu {et~al}\mbox{.}(2011)Sahu, Gurugubelli, Anupama, \&
  Nomoto}]{2011MNRAS.413.2583S}
Sahu D.~K., Gurugubelli U.~K., Anupama G.~C., Nomoto K., 2011, Monthly Notices
  of the Royal Astronomical Society, 413, 2583

\bibitem[{Sauer {et~al}\mbox{.}(2006)Sauer, Mazzali, Deng, Valenti, Nomoto, \&
  Filippenko}]{2006MNRAS.369.1939S}
Sauer D.~N., Mazzali P.~A., Deng J., Valenti S., Nomoto K., Filippenko A.~V.,
  2006, Monthly Notices of the Royal Astronomical Society, 369, 1939

\bibitem[{Schlegel, Finkbeiner \& Davis(1998)Schlegel, Finkbeiner, \&
  Davis}]{1998ApJ...500..525S}
Schlegel D.~J., Finkbeiner D.~P., Davis M., 1998, The Astrophysical Journal,
  500, 525

\bibitem[{Shen {et~al}\mbox{.}(2010)Shen, Kasen, Weinberg, Bildsten, \&
  Scannapieco}]{2010ApJ...715..767S}
Shen K.~J., Kasen D., Weinberg N.~N., Bildsten L., Scannapieco E., 2010, The
  Astrophysical Journal, 715, 767

\bibitem[{Sim {et~al}\mbox{.}(2012)Sim, Fink, Kromer, R\"{o}pke, Ruiter, \&
  Hillebrandt}]{2012MNRAS.420.3003S}
Sim S.~a., Fink M., Kromer M., R\"{o}pke F.~K., Ruiter a.~J., Hillebrandt W.,
  2012, Monthly Notices of the Royal Astronomical Society, 420, 3003

\bibitem[{Stritzinger {et~al}\mbox{.}(2009)Stritzinger, Mazzali, Phillips,
  Immler, Soderberg, Sollerman, Boldt, Braithwaite, Brown, Burns, Contreras,
  Covarrubias, Folatelli, Freedman, Gonz\'{a}lez, Hamuy, Krzeminski, Madore,
  Milne, Morrell, Persson, Roth, Smith, \& Suntzeff}]{2009ApJ...696..713S}
Stritzinger M. {et~al.}, 2009, The Astrophysical Journal, 696, 713

\bibitem[{Sullivan {et~al}\mbox{.}(2011)Sullivan, Kasliwal, Nugent, Howell,
  Thomas, Ofek, Arcavi, Blake, Cooke, Gal-Yam, Hook, Mazzali, Podsiadlowski,
  Quimby, Bildsten, Bloom, Cenko, Kulkarni, Law, \&
  Poznanski}]{2011ApJ...732..118S}
Sullivan M. {et~al.}, 2011, The Astrophysical Journal, 732, 118

\bibitem[{Taubenberger {et~al}\mbox{.}(2008)Taubenberger, Hachinger, Pignata,
  Mazzali, Contreras, Valenti, Pastorello, Elias-Rosa, B\"{a}rnbantner, Barwig,
  Benetti, Dolci, Fliri, Folatelli, Freedman, Gonzalez, Hamuy, Krzeminski,
  Morrell, Navasardyan, Persson, Phillips, Ries, Roth, Suntzeff, Turatto, \&
  Hillebrandt}]{2008MNRAS.385...75T}
Taubenberger S. {et~al.}, 2008, Monthly Notices of the Royal Astronomical
  Society, 385, 75

\bibitem[{Taubenberger {et~al}\mbox{.}(2009)Taubenberger, Valenti, Benetti,
  Cappellaro, {Della Valle}, Elias-Rosa, Hachinger, Hillebrandt, Maeda,
  Mazzali, Pastorello, Patat, Sim, \& Turatto}]{2009MNRAS.397..677T}
Taubenberger S. {et~al.}, 2009, Monthly Notices of the Royal Astronomical Society, 397, 677

\bibitem[{Turatto, Benetti \& Cappellaro(2003)Turatto, Benetti, \&
  Cappellaro}]{2003fthp.conf..200T}
Turatto M., Benetti S., Cappellaro E., 2003, in From Twilight to Highlight: The
  Physics of Supernovae: Proceedings of the ESO/MPA/MPE Workshop Held at
  Garching, INAF, Osservatorio Astronomico di Padova, Vicolo dell'Osservatorio
  5, 35122 Padova, Italia, p. 200

\bibitem[{Turatto {et~al}\mbox{.}(1996)Turatto, Benetti, Cappellaro, Danziger,
  {Della Valle}, Gouiffes, Mazzali, \& Patat}]{1996MNRAS.283....1T}
Turatto M., Benetti S., Cappellaro E., Danziger I.~J., {Della Valle} M.,
  Gouiffes C., Mazzali P.~A., Patat F., 1996, Monthly Notices of the Royal
  Astronomical Society, 283, 1

\bibitem[{Turatto {et~al}\mbox{.}(1998)Turatto, Mazzali, Young, Nomoto,
  Iwamoto, Benetti, Cappellaro, Danziger, de~Mello, Phillips, Suntzeff,
  Clocchiatti, Piemonte, Leibundgut, Covarrubias, Maza, \&
  Sollerman}]{1998ApJ...498L.129T}
Turatto M. {et~al.}, 1998, Astrophysical Journal Letters v.498, 498, L129

\bibitem[{Valenti {et~al}\mbox{.}(2008{\natexlab{a}})Valenti, Benetti,
  Cappellaro, Patat, Mazzali, Turatto, Hurley, Maeda, Gal-Yam, Foley,
  Filippenko, Pastorello, Challis, Frontera, Harutyunyan, Iye, Kawabata,
  Kirshner, Li, Lipkin, Matheson, Nomoto, Ofek, Ohyama, Pian, Poznanski, Salvo,
  Sauer, Schmidt, Soderberg, \& Zampieri}]{2008MNRAS.383.1485V}
Valenti S. {et~al.}, 2008{\natexlab{a}}, Monthly Notices of the Royal
  Astronomical Society, 383, 1485

\bibitem[{Valenti {et~al}\mbox{.}(2008{\natexlab{b}})Valenti, Elias-Rosa,
  Taubenberger, Stanishev, Agnoletto, Sauer, Cappellaro, Pastorello, Benetti,
  Riffeser, Hopp, Navasardyan, Tsvetkov, Lorenzi, Patat, Turatto, Barbon,
  Ciroi, {Di Mille}, Frandsen, Fynbo, Laursen, \&
  Mazzali}]{2008ApJ...673L.155V}
Valenti S. {et~al.}, 2008{\natexlab{b}}, The Astrophysical Journal, 673, L155

\bibitem[{Valenti {et~al}\mbox{.}(2011)Valenti, Fraser, Benetti, Pignata,
  Sollerman, Inserra, Cappellaro, Pastorello, Smartt, Ergon, Botticella,
  Brimacombe, Bufano, Crockett, Eder, Fugazza, Haislip, Hamuy, Harutyunyan,
  Ivarsen, Kankare, Kotak, LaCluyze, Magill, Mattila, Maza, Mazzali, Reichart,
  Taubenberger, Turatto, \& Zampieri}]{2011MNRAS.416.3138V}
Valenti S. {et~al.}, 2011, Monthly Notices of the Royal Astronomical Society, 416, 3138

\bibitem[{Valenti {et~al}\mbox{.}(2009)Valenti, Pastorello, Cappellaro,
  Benetti, Mazzali, Manteca, Taubenberger, Elias-Rosa, Ferrando, Harutyunyan,
  Hentunen, Nissinen, Pian, Turatto, Zampieri, \& Smartt}]{2009Natur.459..674V}
Valenti S. {et~al.}, 2009, Nature, 459, 674

\bibitem[{Valenti {et~al}\mbox{.}(2012)Valenti, Taubenberger, Pastorello,
  Aramyan, Botticella, Fraser, Benetti, Smartt, Cappellaro, Elias-Rosa, Ergon,
  Magill, Magnier, Kotak, Price, Sollerman, Tomasella, Turatto, \&
  Wright}]{2012ApJ...749L..28V}
Valenti S. {et~al.}, 2012, The Astrophysical Journal, 749, L28

\bibitem[{Waldman {et~al}\mbox{.}(2011)Waldman, Sauer, Livne, Perets, Glasner,
  Mazzali, Truran, \& Gal-Yam}]{2011ApJ...738...21W}
Waldman R., Sauer D., Livne E., Perets H., Glasner A., Mazzali P., Truran
  J.~W., Gal-Yam A., 2011, The Astrophysical Journal, 738, 21

\bibitem[{Yaron \& Gal-Yam(2012)}]{2012PASP..124..668Y}
Yaron O., Gal-Yam A., 2012, Publications of the Astronomical Society of the
  Pacific, 124, 668

\bibitem[{Yuan {et~al}\mbox{.}(2013)Yuan, Kobayashi, Schmidt, Podsiadlowski,
  Sim, \& Scalzo}]{Yuan2013}
Yuan F., Kobayashi C., Schmidt B.~P., Podsiadlowski P., Sim S.~A., Scalzo
  R.~A., 2013, Monthly Notices of the Royal Astronomical Society, 432, 1680

\end{thebibliography}

\appendix

\section[]{Tables}

\begin{table*}
 \centering
 \begin{minipage}{160mm}
  \caption{Optical photometry of SN~2012hn (Vega magnitudes in Landolt system)$^a$.}
  \label{tablandolt}
  \footnotesize
  \begin{tabular}{@{}ccccccccc@{}}
  \hline
   Date & JD     & Phase$^{b}$& $U$ & $B$ & $V$ & $R$ & $I$ & Source$^{c}$\\
        & $-$ 2,400,000\\
\hline
  2012-04-13  & 56031.56  & $-2.9$  &  21.561  (200)    &  19.800   (042)    &  17.980  (050)  &  17.310 (027) &   16.790  (020)  &   NTT         \\
  2012-04-15  & 56032.56  & $-1.9$  &    $\cdots$     &  19.944  (141)    &  17.903  (054)  &  17.286 (066) &   16.751  (032)  &   PROMPT 5 \\
  2012-04-17  & 56034.56  &   0.0   &    $\cdots$     &   $\cdots$        &  17.975  (065)  &  17.272 (079) &   16.686  (053)  &   PROMPT 5 \\
  2012-04-18  & 56035.50  &   1.0   &    $\cdots$     &   $\cdots$        &  18.074  (039)  &  17.294 (047) &   16.736  (047)  &   PROMPT 5 \\
  2012-04-21  & 56038.55  &   4.0   &    $\cdots$     &  20.443  (268)    &  18.277  (073)  &  17.393 (089) &   16.741  (051)  &   PROMPT 5 \\
  2012-04-20  & 56038.58  &   4.1   &    $\cdots$     &  20.490  (037)    &  18.206  (027)  &  17.450 (021) &   16.720  (018)  &   NTT          \\ 
  2012-04-22  & 56040.49  &   6.0   &    $\cdots$     &  20.620  (237)    &  18.397  (083)  &  17.570 (102) &   16.891  (053)  &   PROMPT 5 \\
  2012-04-23  & 56040.53  &   6.0   &  22.282  (200)    &  20.800  (054)    &  18.320  (030)  &  17.540 (025) &   16.830  (031)  &   NTT          \\
  2012-04-29  & 56047.56  &  13.1   &    $\cdots$     &  21.510  (064)    &  19.050  (072)  &  18.080 (027) &   17.290  (036)  &   NTT           \\
  2012-04-30  & 56048.48  &  14.0   &    $\cdots$     &  21.530  (078)    &  19.020  (050)  &  18.160 (043) &   17.320  (039)  &   NTT           \\
  2012-05-19  & 56066.98  &  32.5   &    $\cdots$     &    $\cdots$       &  19.936  (045)  &  18.952 (039) &  $\cdots$        &   TRAPPIST    \\
  2012-05-19  & 56066.98  &  32.5   &    $\cdots$     &    $\cdots$       &  $\cdots$       &   $\cdots$    &   18.026  (031)  &   PROMPT5    \\
  2012-05-19  & 56067.50  &  33.0   &    $\cdots$     &  22.210  (032)    &  19.951  (031)  &   $\cdots$    &  $\cdots$        &   DUPONT     \\
  2012-08-10  & 56149.91  & 115.4   &    $\cdots$     &    $\cdots$       &  $\cdots$       &  21.029 (035) &  $\cdots$        &   NTT            \\
  2012-08-25  & 56164.87  & 130.4   &    $\cdots$     &    $\cdots$       &  22.683  (160)  &  21.354 (023) &  20.427  (020)   &   NTT             \\
  2012-09-09  & 56180.89  & 146.4   &    $\cdots$     &    $\cdots$       &  23.431  (560)  &  21.716 (286) &  20.932  (250)   &   NTT             \\
  2012-10-16  & 56217.81  & 183.3   &    $\cdots$     &    $\cdots$       &  23.762  (270)  &  22.514 (256) &  21.721  (174)   &   NTT             \\
  2012-11-15  & 56246.73  & 212.2   &    $\cdots$     &    $\cdots$       &  $\cdots$       &  22.838 (083) &  22.005  (081)   &   NTT             \\
  2012-12-04  & 56265.82  & 231.3   &    $\cdots$     &    $\cdots$       &  $\cdots$       &  23.020 (204) &  $\cdots$        &   NTT             \\
  2012-12-06  & 56267.81  & 233.3   &    $\cdots$     &    $\cdots$       &  $\cdots$       &  23.470 (217) &  22.607  (145)   &   NTT             \\
\hline
 \end{tabular}\\
$^a$\,The errors are computed taking into account 
both the uncertainty of the PSF fitting of the SN magnitude and 
the uncertainty due to the background contamination 
(computed by an artificial-star experiment).
$^b$\,Relative to the $R$-band maximum (JD = 2,456,034.5). 
$^c$\,PROMPT $=$ PROMPT Telescopes and CCD camera Alta U47UV E2V CCD47-10; pixel scale $=$ 0.590 arcsec pixel$^{-1}$.
$~$ NTT $=$ New Technology Telescope and EFOSC2; pixel scale $=$ 0.24 arcsec pixel$^{-1}$. 
$~$ TRAPPIST $=$ TRAnsiting Planets and PlanetesImals Small Telescope and FLI CCD; pixel scale $=$ 0.64 arcsec pixel$^{-1}$. 
$~$ DUPONT $=$ 2.5-m du Pont telescope and SITe2k; pixel scale $=$ 0.259 arcsec pixel$^{-1}$.
\end{minipage}
\end{table*}

\begin{table*}
 \centering
 \begin{minipage}{160mm}
  \caption{Optical photometry of SN~2012hn (AB magnitudes in Sloan system)$^a$.}
  \label{tabsloan}
  \begin{tabular}{@{}cccccccc@{}}
  \hline
  Date & JD  & Phase$^{b}$ & $g$ & $r$ & $i$ & $z$ & Source$^{c}$\\
        & $-$ 2,400,000\\
\hline
   2012-04-14  & 56031.51  & $-3.0$ &  18.367 (069)  &   $\cdots$     &     $\cdots$    &   $\cdots$     &    NTT  \\
   2012-04-14  & 56031.51  & $-3.0$ &  18.398 (138)  &   $\cdots$     &     $\cdots$    &   $\cdots$     &    NTT   \\
   2012-04-15  & 56032.59  & $-1.9$ &    $\cdots$    &  17.667 (200)  &     $\cdots$    &   $\cdots$     &    PROMPT 5   \\
   2012-04-20  & 56037.53  &  3.0   &  18.821 (085)  &  17.614 (092)  &  17.198 (061)   &  16.972 (039)  &    PROMPT 5/3  \\
   2012-04-21  & 56038.52  &  4.0   &  18.882 (069)  &   $\cdots$     &     $\cdots$    &   $\cdots$     &    NTT  \\
   2012-04-21  & 56038.53  &  4.0   &  18.852 (066)  &   $\cdots$     &     $\cdots$    &   $\cdots$     &    NTT  \\
   2012-04-24  & 56041.48  &  7.0   &  19.276 (157)  &  17.760 (170)  &     $\cdots$    &   $\cdots$     &    PROMPT 5/3 \\
   2012-04-30  & 56048.48  & 14.0   &  19.907 (110)  &  18.425 (071)  &   17.852 (025)  &  17.485 (035)  &    NTT   \\
\hline	 		 
\end{tabular}\\
$^a$\,The errors are computed taking into account 
both the uncertainty of the PSF fitting of the SN magnitude and 
the uncertainty due to the background contamination 
(computed by an artificial-star experiment).
$^b$\,Relative to the $R$-band maximum (JD = 2,456,034.5).
$^c$\,PROMPT $=$ PROMPT Telescopes and CCD camera Alta U47UV E2V CCD47-10; pixel scale $=$ 0.590 arcsec pixel$^{-1}$.
$~$ NTT $=$ New Technology Telescope and EFOSC2; pixel scale $=$ 0.24 arcsec pixel$^{-1}$. 
\end{minipage}
\end{table*}

\begin{table*}
  \caption{Optical photometry of SN~2012hn reference stars (Vega magnitudes in  Landolt system)$^a$.}
  \label{tabseqstar1}
  \footnotesize
  \begin{tabular}{@{}cccccc@{}}
  \hline
Id & $U$ & $B$ & $V$ & $R$ & $I$ \\
\hline
  1   &  16.909   (068)  &    16.346  (021) & 15.454  (017) & 14.941  (020) &  14.453  (017)  \\
  2   &  17.390   (033)  &    17.400  (053) & 16.783  (065) & 16.424  (074) &  16.000  (073)  \\
  3   &  17.565   (043)  &    17.075  (059) & 16.207  (089) & 15.724  (083) &  15.231  (091)  \\
  4   &  20.813   (026)  &    19.532  (068) & 17.905  (163) & 16.420  (107) &  14.637  (085)  \\
  5   &  16.323   (025)  &    16.326  (060) & 15.781  (083) & 15.471  (086) &  15.109  (059)  \\
  6   &  17.873   (074)  &    17.921  (029) & 17.461  (053) & 17.161  (053) &  16.833  (026)  \\
  7   &  17.258   (051)  &    16.907  (005) & 16.064  (020) & 15.593  (031) &  15.110  (024)  \\
  8   &  15.594   (035)  &    15.613  (011) & 15.002  (024) & 14.639  (029) &  14.275  (045)  \\
  9   &  19.894   (075)  &    18.684  (046) & 17.210  (029) & 16.248  (023) &  15.160  (032)  \\
 10   &  16.912   (046)  &    16.990  (035) & 16.467  (070) & 16.135  (063) &  15.777  (067)  \\
\hline
\end{tabular}\\
$^a$The uncertainties are the standard deviation of the mean
 of the selected measurements.
\end{table*}

\begin{table*}
  \caption{Optical photometry of SN~2012hn reference stars (AB magnitudes in Sloan system)$^a$.}
  \label{tabseqstar2}
  \begin{tabular}{@{}ccccc@{}}
  \hline
Id & $g$ & $r$ & $i$ & $z$ \\
\hline 
   1  &     15.74    (07) &  15.15  (06) & 14.89  (04)  &  14.71   (02) \\
   2  &     16.86    (07) &  16.48  (06) & 16.31  (04)  &  16.27   (02) \\
   3  &     16.41    (06) &  15.79  (06) & 15.52  (04)  &  15.41   (02) \\
   4  &     18.36    (07) &  16.85  (06) & 15.30  (03)  &  14.39   (02) \\
   5  &     15.82    (06) &  15.49  (05) & 15.40  (03)  &  15.38   (02) \\
   6  &     17.52    (07) &  17.26  (06) & 17.22  (04)  &  17.19   (02) \\
   7  &     16.28    (06) &  15.75  (05) & 15.51  (04)  &  15.42   (02) \\
   8  &     15.12    (06) &  14.79  (05) & 14.64  (03)  &  14.59   (02) \\
   9  &     17.89    (07) &  16.66  (06) & 15.78  (04)  &  15.32   (02) \\
  10  &     16.51   (07) &  16.17  (06) & 16.06  (04)  &  16.05   (02) \\
\hline
\end{tabular}\\
$^a$The uncertainties are the standard deviation of the mean
 of the selected measurements.
\end{table*}

\end{document}